\documentclass[nofootinbib,aps,prx,twocolumn,superscriptaddress,showpacs,amsmath,amssymb,longbibliography,10pt,floatfix]{revtex4-2}
\usepackage{graphicx}
\usepackage[breaklinks,colorlinks,linkcolor=blue,citecolor=blue,urlcolor=blue]{hyperref}

\usepackage[utf8]{inputenc}
\usepackage[english]{babel}
\usepackage{amsmath}
\usepackage{xcolor}

\definecolor{nred}{RGB}{224,0,0}
\definecolor{nblue}{RGB}{28,130,185}

\usepackage[normalem]{ulem} % Težave v bibliografiji, kjer se naslovi knjig ne zlomijo preko vrstice (npr. pri citiranju horvatic01 ali shirane06), se odpravi z opcijo "normalem" za ta package. Ta opcija je tako NUJNA, da je bibliografija prava!

\usepackage[nameinlink,capitalize]{cleveref}
\crefname{section}{Sec.}{Secs.}
\crefname{pluralfigure}{Figs.}{Figs.}
\crefname{pluralequation}{Eqs.}{Eqs.}
\creflabelformat{pluralequation}{#2(#1)#3}
\usepackage{xspace}
\usepackage{chemformula}
\newcommand{\herbFormula}{\ch{ZnCu3(OH)6Cl2}\xspace} %\newcommand{\herbFormula}{ZnCu$_3$(OH)$_6$Cl$_2$\xspace}
\newcommand{\ycuFormula}{\ch{YCu3(OH)6Cl3}\xspace} %\newcommand{\ycuFormula}{YCu$_3$(OH)$_6$Cl$_3$\xspace}

\usepackage{siunitx}  % For the \SI{}{} command
\usepackage{braket}   % For \bra, \ket, \braket (MUCH better typesetting than \langle ... \rangle !)
\AtBeginDocument{\usepackage{booktabs}} % For "\toprule, \midrule, \bottomrule" in tables ("http://en.wikibooks.org/wiki/LaTeX/Tables#Professional_tables", "http://www.inf.ethz.ch/personal/markusp/teaching/guides/guide-tables.pdf", "https://tex.stackexchange.com/questions/222697/how-to-use-booktab-or-type-this-table-in-revtex4-revtex4-1")

\DeclareMathOperator{\tr}{tr}
\newcommand*\dd{\mathop{}\!\mathrm{d}}

\definecolor{darkgreen}{RGB}{0,128,0}
\definecolor{darkblue}{RGB}{0,0,128}
\definecolor{nblue2}{RGB}{24,118,178}
\definecolor{nyellow}{RGB}{205,116,0} %\definecolor{nyellow}{RGB}{224,153,0}

\begin{document}

%\title{Dynamical spin correlations in kagome-lattice antiferromagnetic model and materials}
%\title{Dynamical spin correlations on the kagome lattice}  
%\title{Finite-temperature dynamical spin correlations on the kagome lattice}  
%\title{$T$-dependent dynamical spin correlations on the kagome lattice}  
\title{Dynamical spin correlations of the kagome antiferromagnet}  
\author{P. Prelov\v{s}ek}
\affiliation{Jo\v{z}ef Stefan Institute, Jamova c.~39, SI-1000 Ljubljana, Slovenia}
\affiliation{Faculty of Mathematics and Physics, University of Ljubljana, Jadranska u.~19, SI-1000 Ljubljana, Slovenia}
\author{M. Gomil\v{s}ek}
\affiliation{Jo\v{z}ef Stefan Institute, Jamova c.~39, SI-1000 Ljubljana, Slovenia}
\author{T. Arh}
\affiliation{Jo\v{z}ef Stefan Institute, Jamova c.~39, SI-1000 Ljubljana, Slovenia}
\affiliation{Faculty of Mathematics and Physics, University of Ljubljana, Jadranska u.~19, SI-1000 Ljubljana, Slovenia}
\author{A. Zorko}
\email{andrej.zorko@ijs.si}
\affiliation{Jo\v{z}ef Stefan Institute, Jamova c.~39, SI-1000 Ljubljana, Slovenia}
\affiliation{Faculty of Mathematics and Physics, University of Ljubljana, Jadranska u.~19, SI-1000 Ljubljana, Slovenia}

\begin{abstract}
Temperature-dependent dynamical spin correlations, which can be readily accessed via a variety of experimental techniques, hold the potential of offering a unique fingerprint of quantum spin liquids and other intriguing dynamical states.
In this work we present an in-depth study of the temperature-dependent dynamical spin structure factor $S({\bf q}, \omega)$ of the 
antiferromagnetic (AFM) Heisenberg spin-1/2 model on the kagome lattice with additional Dzyaloshinskii--Moriya (DM) interactions.
Using the finite-temperature Lanczos method on lattices with up to $N = 30$ sites we find that even without DM interactions, chiral low-energy spin fluctuations of the $\ang{120}$ AFM order parameter dominate the dynamical response.
This leads to a nontrivial frequency dependence of $S({\bf q}, \omega)$ and the appearance of a pronounced low-frequency mode at the M point of the extended Brillouin zone.
Adding an out-of-plane DM interactions $D^z$ gives rise to an anisotropic dynamical response, a softening of in-plane spin fluctuations, and, ultimately, the onset of a coplanar AFM ground-state order at $D^z > 0.1 J$.
Our results are in very good agreement with existing inelastic neutron scattering and temperature-dependent NMR spin-lattice relaxation rate ($1/T_1$) data on the paradigmatic kagome AFM herbertsmithite, where the effect of its small $D^z$ on the dynamical spin correlations is shown to be rather small, as well as with $1/T_1$ data on the novel kagome AFM \ycuFormula, where its substantial $D^z \approx 0.25 J$ interaction is found to strongly affect the spin dynamics.

\end{abstract}
%================================================================================
\maketitle

%================================================================================
\section {Introduction} 

The antiferromagnetic (AFM) Heisenberg spin-1/2 model on the kagome lattice (KLHM) is one of the most intensively studied quantum spin models, owing to its unique ground state (GS) and low-$T$ properties \cite{lee08,balents10,savary17,broholm20}. 
Various theoretical and numerical investigations have established KLHM as the most promising candidate amongst isotropic spin models to feature a quantum spin liquid (SL) GS, where the absence of low-$T$ long-range order is accompanied by strong quantum entanglement between constituent spins.
However, the nature of the SL GS, including the presence of either a finite \cite{mila98,waldtmann98,singh07,singh08,yan08,lauchli11,iqbal11,depenbrock12,schnack18,lauchli19,prelovsek20} or a vanishing \cite{ran07,iqbal13,xie14,iqbal14,he17,liao17} energy gap $\Delta_t$ to spin-triplet excitations, remains controversial.
Properties of the KLHM at finite temperatures may provide important insights into this long-standing issue.

Thermodynamic quantities such as the uniform susceptibility $\chi_0(T)$, magnetic specific heat $c(T)$, and the related entropy density $s(T)$ of the KLHM
have previously been studied by high-$T$ series expansion \cite{misguich07,bernu20}, via numerical linked cluster methods \cite{rigol071,rigol072},
and more recently with the finite-temperature Lanczos method (FTLM) \cite{schnack18,prelovsek20,arh20} on finite spin systems with up to $N=42$ sites. 
Apart from an evidence of finite spin triplet gap $\Delta_t >0$, FTLM results indicate that there is a substantial remnant entropy $s(T)>0$ at very low $T$, which is a signature of a large density of low-energy singlet excitations
with a (nearly) vanishing spin singlet energy gap $\Delta_s \ll \Delta_t$.
The static (equal-time) spin correlation function $S^{\alpha \alpha}({\bf q})$ has also been studied both at $T = 0$ \cite{iqbal13} and at finite temperatures \cite{shimokawa16,sherman18}. 
However, dynamical spin properties of the KLHM, 
in particular the dynamical spin structure factor (DSF) $S^{\alpha \alpha}({\bf q},\omega)$, 
are theoretically poorly understood even though the temperature-dependent DSF is potentially a unique fingerprint of SL states, and is experimentally directly accessible via inelastic neutron-scattering (INS) and nuclear magnetic resonance (NMR) relaxation \cite{sherman16}.
Because of its fundamental importance various analytical concepts and methods \cite{hao10,punk2014,zhu2019}, as well as numerical approaches \cite{shimokawa15,sherman18}, have been employed to study it, though they have mostly led to inconclusive results.

One reason for the theoretical difficulties lies in the large density of low-energy spin-singlet states of the KLHM \cite{prelovsek20}, which implies that a meaningful evaluation of the DSF would require a more challenging finite-temperature instead of GS treatment.
Another reason is that the DSF of a SL, like the one in KLHM, is usually (implicitly) assumed to be rather featureless due to the fractionalization of spin excitations. 
We show that the KLHM DSF instead has some quite pronounced spectral features.

On the experimental front, investigations of the KLHM have been boosted in the last couple of decades by the discovery of several promising kagome-lattice (KL) materials exhibiting SL properties at low temperatures. 
The most prominent example is herbertsmithite, \herbFormula \cite{shores05,mendels10,norman16}, 
where the availability of single crystals allows for full access to the DSF $S^{\alpha \alpha}({\bf q},\omega)$ \cite{han12,fu15,khuntia20}. 
While several other KL materials have been discovered in recent years \cite{hiroi01,fak12,li14,gomilsek16,feng17,zorko191}, we will mostly focus on 
the recently synthesized \cite{sun16} and investigated \cite{barthelemy19,zorko191,zorko192,arh20} compound
\ycuFormula, which has the distinct advantage of having a structurally-perfect kagome lattice without any substitutional disorder, in contrast to most other KL materials including herbertsmithite \cite{mendels10,norman16,zorko17}.
Besides potential imperfections the relation of KL materials to the ideal KLHM is often further complicated by additional Dzyaloshinskii--Moriya (DM) interactions, which are usually allowed in these systems as most lack local inversion symmetry on superexchange bonds $J$ between nearest-neighbor magnetic ions.
While weak DM interactions are expected to lead to mostly quantitative corrections of observables at low $T$ \cite{rigol072,bernu20}, as in the case of herbertsmithite \cite{zorko08,el10}, strong DM interactions can lead to a quantum phase transition from a SL to a long-range ordered (LRO) GS \cite{cepas08,elhajal02,zorko13}, as in the case of \ycuFormula where an out-of-plane $D^z \approx 0.25 J$ induces chiral $\ang{120}$ AFM LRO \cite{zorko192,arh20}. 
The addition of DM interactions to the KLHM is therefore crucial for explaining the observed properties of many KL materials,
especially low-$T$ ordered ones like \ycuFormula.

In this paper we present a comprehensive numerical study of the DSF $S^{\alpha \alpha}({\bf q},\omega)$ of the KLHM with additional out-of-plane DM interactions $D = D^z$ at finite temperatures. To this end we employ the FTLM on systems with up to $N = 30$ sites under periodic boundary conditions. This method is particularly suitable for frustrated spin systems (and in general strongly-correlated systems) that do not possess long-range correlations down to $T \ll J$, which allows us to obtain static and dynamical properties of macroscopic validity 
down to temperatures many times lower \cite{prelovsek20,arh20} 
than in systems with GS LRO  \cite{jaklic00, prelovsek20}.
In contrast to previous investigations of the KLHM DSF \cite{sherman18} we find that it is in fact not featureless.
Even at $D = 0$ we find particularly pronounced low-energy chiral $\ang{120}$ AFM fluctuations corresponding to the wavevector $q = 0$ in the reduced Brillouin zone (BZ), or, equivalently, the M point of the extended BZ. Furthermore, we find that the low-$T$, low-energy DSF of the KLHM seems to be governed by a finite spin triplet gap $\Delta_t > 0$.
Adding finite DM interactions $D > 0$ results in an anisotropic DSF and a softening of the in-plane spin triplet gap $\Delta^x_t$ that ultimately leads to GS LRO for $D > D_c \approx 0.1 J$.
The calculated DSF is also used to evaluate temperature-dependent local spin fluctuation (LSF) spectra $S_L^{\alpha \alpha}(\omega)$, which are directly related to experimental NMR spin-lattice relaxation rates $1/T_1$.
Finally, the obtained numerical DSF and LSF results are compared with experimental INS \cite{han12} and NMR results \cite{fu15} on herbertsmithite and on the impurity-free \ycuFormula \cite{arh20b}.

%================================================================================
\section{Model, numerical method and considered quantities}\label{sec2}

\begin{figure}[!t]
\centering
\includegraphics[width=0.8\columnwidth]{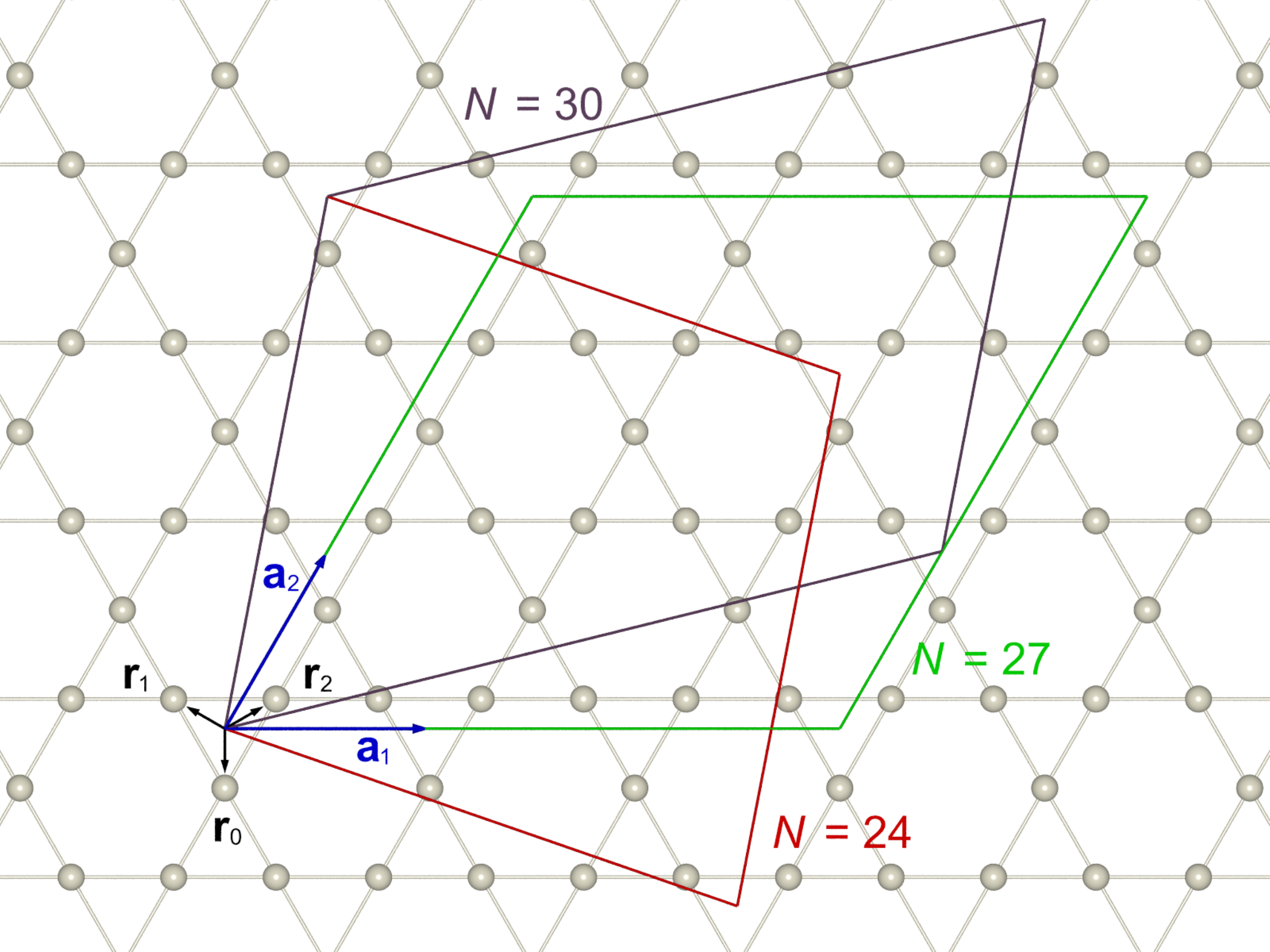}
\caption{Finite size kagome lattices with $N=24$, $27$ and $30$ sites used in our FTLM calculations. The primitive vectors of the underlying hexagonal Bravais lattice are denoted by ${\bf a}_1$ and ${\bf a}_2$, while the three basis vectors of the kagome lattice are denoted by ${\bf r}_0$, ${\bf r}_1$, and ${\bf r}_2$.}
\label{fig1}
\end{figure}

We consider the KLHM with AFM isotropic Heisenberg nearest-neighbor exchange interactions $J$ between $S=1/2$ 
spins on a KL with additional out-of-plane DM interactions $D$,
\begin{equation}
\mathcal{H} =  \sum_{\langle ij \rangle}  \Bigl[ J \, {\bf S}_i \cdot {\bf S}_j + D ({\bf S}_i \times {\bf S}_j)^z \Bigr] , 
\label{eq_hm}
\end{equation}
where $\langle ij \rangle$ is a sum over nearest-neighbor spin pairs and the spins in the DM term appear in the clockwise direction around each lattice triangle (see \cref{fig1}).
Except in \cref{sec5} where we compare our numerical results with experiment, we use $\hbar = k_B = 1$ units as well as $J=1$. 
All energies, frequencies and temperatures are thus implicitly given relative to $J$. In KL materials the DM interaction has the general
form ${\bf D}_{ij} \cdot ({\bf S}_i \times {\bf S}_j)$ where ${\bf D}_{ij}$ is a vector
with an out-of-plane component $D^z_{ij}$ and an in-plane component $D^p_{ij}$.
In this paper we consider only the effect of a non-zero $D^z_{ij} = D$ for three reasons.
Firstly, an in-plane $D^p_{ij}$ is symmetry-allowed only when the kagome plane is not also a crystallographic mirror plane \cite{elhajal02}, and is thus present less often.
Secondly, the effect of $D^p_{ij} \neq 0$ appears to be weaker and qualitatively less important than that of $D^z_{ij} \neq 0$ in the KLHM, as confirmed both theoretically and experimentally \cite{rigol072,cepas08,bernu20,arh20}. 
And thirdly, as a practical benefit, when $D^p_{ij} = 0$ the hamiltonian $\mathcal{H}$ remains uniaxially symmetric about the $z$ axis, conserving the $z$ component of total magnetization $S^z_\mathrm{tot} = \sum_i S^z_i$, which significantly reduces the dimensionality of invariant Hilbert subspaces, and hence the memory requirements, of the FTLM.

The standard definition of the DSF is
\begin{equation}
S^{\alpha\beta}({\bf q},\omega) = \frac{1}{2\pi} \int_{-\infty}^\infty \dd t~ e^{i \omega t} \braket{S^\alpha_{-{\bf q}}(t) S^\beta_{\bf q}(0)} ,
\label{eq_dsfs}
\end{equation}
where $\braket{\dots}$ denotes the canonical thermal average, $\alpha$ and $\beta$ are components of ${\bf q}$-space spin operators ${\bf S}_{\bf q} = (1/\sqrt{N}) \sum_i e^{i {\bf q} \cdot {\bf R}_i} {\bf S}_i$ defined via the positions ${\bf R}_i$ of spins in the KL, and $N$ is the total number of KL sites. 
As the KL is formed of three basis vectors ${\bf r}_k$ ($k = 0, 1, 2$) on an underlying hexagonal 
Bravais lattice of down-pointing triangle centers $\widetilde{\bf R}_n$ (see \cref{fig1}) one has ${\bf R}_i = \widetilde{\bf R}_n + {\bf r}_k$ where $i \equiv (n, k)$.
Due to the three KL basis vectors the DSF is only ${\bf q}$-periodic over an extended BZ that is 4-times larger than the reduced BZ of the underlying hexagonal Bravais lattice (see \cref{fig5}).

A more insightful definition of the DSF for the KL, 
which explicitly takes into account its threefold rotational symmetry, 
and which is beneficial both numerically as well as for theoretical understanding,
instead involves chiral spin operators in ${\bf q}$-space,
\begin{equation}
\widetilde{\bf S}_{c{\bf q}} = \frac{1}{\sqrt{N}} \sum_n e^{i {\bf q} \cdot \widetilde{\bf R}_n} \left[ {\bf S}_{(n,0)} + \zeta^c {\bf S}_{(n,1)} + \zeta^{-c} {\bf S}_{(n,2)} \right] , 
\label{eq_sqc}
\end{equation}
where $n$ in $(n,k)$ runs over all down-pointing triangles of the KL, $k$ runs over the three spins inside these triangles, 
$\zeta = e^{2 \pi i/3}$, and $c=-1,0,1$ denotes the vector spin chirality of the KL triangles. 
Note that the standard $\ang{120}$ AFM LRO on the KL involves only the chiral spin operators with $c = \pm 1$, 
while ferromagnetic LRO involves the $c=0$ chiral spin operators.
Using \cref{eq_sqc} we define the (diagonal) chiral DSF on the KL as
\begin{equation}
\widetilde{S}_c^{\alpha\beta}({\bf q},\omega) = \frac{1}{2\pi} \int_{-\infty}^\infty \dd t~ e^{i \omega t} \braket{\widetilde{S}^{\alpha \dagger}_{c{\bf q}}(t) \widetilde{S}^\beta_{c{\bf q}}(0)} , 
\label{eq_chiral_dsf}
\end{equation}
which is ${\bf q}$-periodic over the reduced BZ, not just over the larger extended BZ of the standard DSF of \cref{eq_dsfs} (see \cref{fig5}). Note that at a generic ${\bf q}$ one could also expect non-vanishing off-diagonal terms $\braket{\widetilde{S}^{\alpha \dagger}_{c{\bf q}}(t) \widetilde{S}^\beta_{c'{\bf q}}(0)}$ with $c \neq c'$. 
Nevertheless, these terms are expected to be less important than the diagonal ones, and are much more difficult to handle within the FTLM, so we neglect them. 
We can then express the standard DSF of \cref{eq_dsfs} from the chiral DSF of \cref{eq_chiral_dsf} as
\begin{equation}
\begin{split}
S^{\alpha\alpha}({\bf q},\omega) &= \sum_{c=-1}^1 |\xi_c({\bf q})|^2 \widetilde{S}_c^{\alpha\alpha}({\bf q},\omega) , \\
\xi_c({\bf q}) &= \frac{1}{3} \sum_{k=0}^2 e^{i {\bf q} \cdot {\bf r}_k} \zeta^{-c k} ,  \label{eq_dsf_from_chiral_dsf}
\end{split}
\end{equation}
where  $\widetilde{S}_c^{\alpha\beta}({\bf q},\omega) = 0$ for $\alpha \neq \beta$ and $S^{xx}({\bf q},\omega) = S^{yy}({\bf q},\omega)$ since $S^z_\mathrm{tot}$ is a conserved quantity.

We evaluate the chiral DSF at $T>0$ using the FTLM, introduced in Refs.~\cite{jaklic94,jaklic00} and used in numerous studies of static and dynamical properties of various correlated systems \cite{prelovsek13}. 
In the case of the KLHM, the FTLM has previously been employed only for the calculation of thermodynamic quantities, such as the uniform susceptibility $\chi_0(T)$, entropy density $s(T)$ and specific heat $c(T)$ \cite{schnack18,prelovsek20,arh20}, that involve only the conserved quantities of energy and total magnetization $S^z_\mathrm{tot}$.
In contrast, the evaluation of the chiral DSF (here given in the Lehmann representation) is more involved,
\begin{equation}
\widetilde{S}_c^{\alpha\alpha}({\bf q},\omega) = \frac{1}{Z} \sum_n e^{-\epsilon_n/T} \braket{\psi_n | \widetilde{S}^{\alpha \dagger}_{c {\bf q}} \delta (\omega + \epsilon_n - H ) \widetilde{S}^\alpha_{c{\bf q}} | \psi_n} ,
\label{eq_chiral_dsf_lehmann}
\end{equation}
where $Z = \sum_n e^{-\epsilon_n/T}$ is the canonical partition function, $\ket{\psi_n}$ are eigenfunctions of $\mathcal{H}$ and $\epsilon_n$ are their eigenenergies.
As the chiral DSF already takes into account both translation symmetry and the conservation of $S^z_\mathrm{tot}$, the needed Hilbert subspaces remain the same as for static quantities; e.g., the largest subspace for $N=30$ sites contains $N_\mathrm{st} \sim 10^7$ states.
In FTLM we replace $\sum_n$ over all eigenfunctions with a trace over $R >1$ random initial wavefunctions $\ket{r}$ and the expectation value with a double sum over the emerging Lanczos (eigen)functions $\ket{\phi^r_i}$, $\ket{\widetilde{\phi}^r_j}$ in different ${\bf q}$ wavevector sectors \cite{jaklic94,jaklic00,prelovsek13}, 
with $i,j \leq N_L$ where $N_L$ is the number of performed Lanczos steps.  
This requires additional storage of $2 N_L$ wavefunctions 
meaning that the total memory requirements for the dynamical FTLM are $\mathcal{O}(N_L N_\mathrm{st})$. 
To achieve satisfactory $\omega$ resolution in the DSF $N_L > 100$ is typically required.

In the following we evaluate the chiral DSF on several finite-sized lattices with $N=24$, $27$ and $30$ sites (\cref{fig1}).
While the $N=24$ and $30$ lattices break the rotational symmetry of the infinite KL the $N=27$ lattice preserves it, but is less convenient because of its $S_\mathrm{tot} = 1/2$ GS, whereas the infinite KLHM should have a $S_\mathrm{tot} = 0$ GS \cite{lee08,balents10,savary17,broholm20}.
While for $N=24$ and $27$ we can afford $N_L \sim 200$ and $R>10$, most of the present results are for $N=30$ sites where we used $N_L = 120$ and $R = 3$ within each symmetry sector.
We note that the main criterion for (even macroscopic) validity of FTLM results (in the given model and system size) is that the modified thermodynamic sum $\widetilde Z(T) =R ~\mathrm{Tr}[ \exp( -(H-E_0)/T)] > \widetilde Z(T_\mathrm{fs}) \gg 1$ \cite{jaklic94,jaklic00}, where $E_0$ is the ground-state energy and trace $\mathrm{Tr}$ involves the sum over all wave vector and $S^z_\mathrm{tot}$ sectors. 
Due to very large density of 
low-lying states in SL systems (and directly related large  entropy even at low $T$), even modest $R=3$ is enough to reach valid 
results
down to temperatures $T > T_\mathrm{fs} \sim 0.1 J + D$ \cite{prelovsek20,arh20} below which they are limited by finite-size effects, i.e.,
by the onset of longer-range correlations for $D>0$.  

Finally, while $S^{\alpha\alpha}({\bf q},\omega)$ contains all of the dynamical information, it is also useful to extract the equal-time spin correlation function
$S^{\alpha\alpha}({\bf q})$ and the d.c. spin susceptibility $\chi_0^{\alpha\alpha}({\bf q})$, 
defined from the DSF as
\begin{equation}
\begin{split}
S^{\alpha\alpha}({\bf q}) &= \int_{-\infty}^\infty \dd\omega~ S^{\alpha\alpha}({\bf q},\omega)  = 
\braket{S^\alpha_{-{\bf q}} S^\alpha_{\bf q}} , \\
\chi^{\alpha\alpha}_0({\bf q}) &= \mathcal{P} \int_{-\infty}^\infty \dd\omega~ \frac{1- e^{-\omega/T}}{\omega} S^{\alpha\alpha}({\bf q},\omega) ,
\label[pluralequation]{eqs_quantities}
\end{split}
\end{equation}
where $\mathcal{P}$ denotes the Cauchy principal value.
Note that $\omega^\alpha( {\bf q}) = \sqrt{S^{\alpha\alpha}({\bf q})/ \chi^{\alpha\alpha}_0({\bf q})  }$ can be interpreted as the characteristic spin-fluctuation frequency at a given ${\bf q}$ and for a given direction $\alpha$.

%================================================================================
\section{Heisenberg model on kagome lattice}\label{sec3}

In this section we consider the $D=0$ KLHM. Since this model is isotropic in spin space it has an isotropic chiral DSF $\widetilde{S}_c^{\alpha\alpha}({\bf q},\omega) = \widetilde{S}_c({\bf q},\omega)$ and hence also isotropic derived quantities in \cref{eqs_quantities}.
In the following we present numerical results for the standard choice $\alpha=z$, which is numerically less costly to evaluate since the relevant operators are diagonal in the $S^z_\mathrm{tot}$ basis.

\subsection{Dynamical spin structure factor}\label{sec3a}

%--------------------------------------------------------------------------------
\begin{figure}[!t]
\centering
\includegraphics[width=0.8\columnwidth]{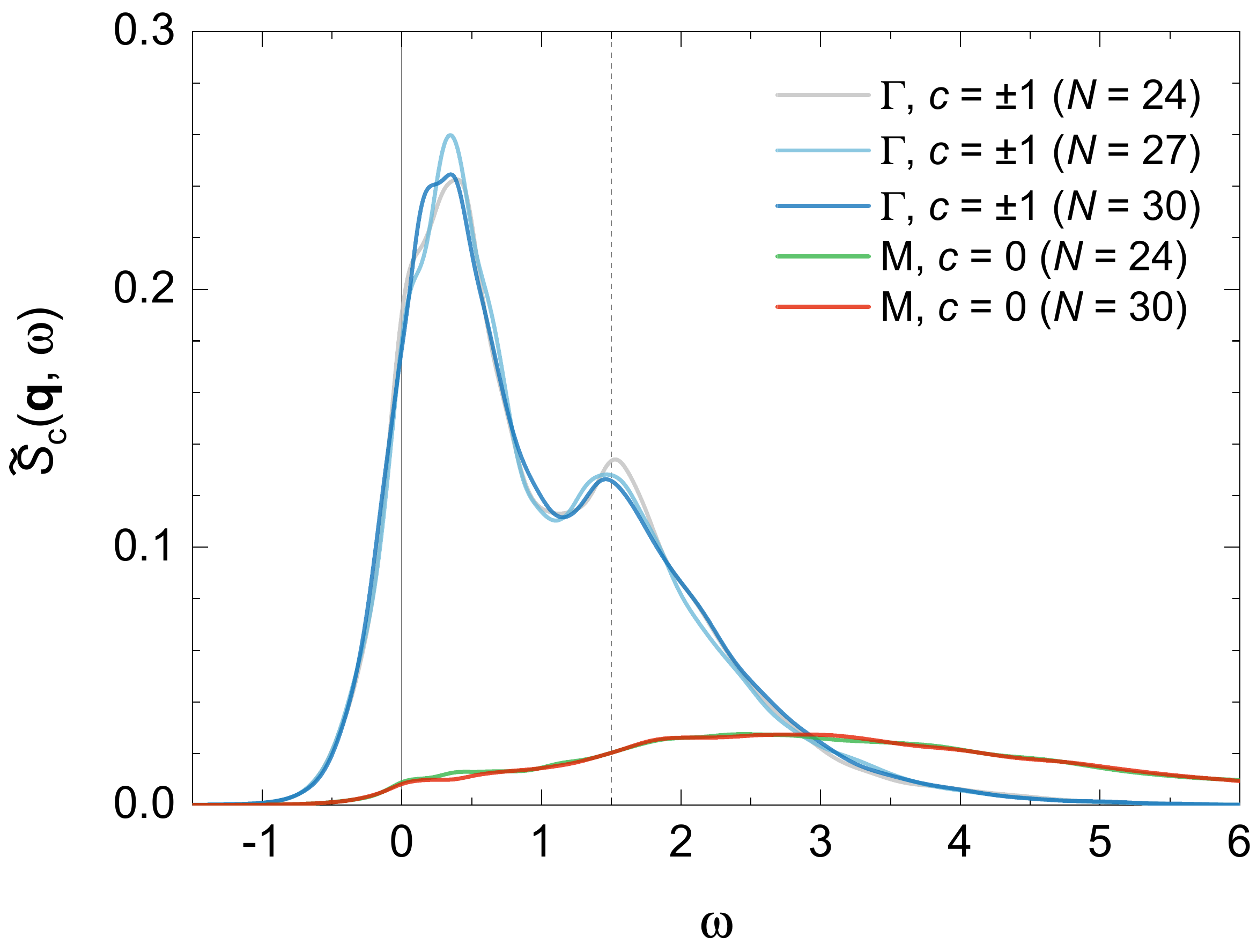}
\caption{ Chiral DSF's $\widetilde{S}_{\pm 1}(q=0, \omega)$  and $\widetilde{S}_0({\bf q} = \mathrm{M}, \omega)$ at $T = 0.2$ calculated using the FTLM 
for different lattice sizes $N=24$--$30$ (\cref{fig1}). 
Note that the $N=27$ lattice does not contain the M point of the reduced BZ. 
The vertical dotted line at $\omega = 1.5$ corresponds to triplet excitations within an isolated Heisenberg spin triangle.}
\label{fig2}
\end{figure}
%--------------------------------------------------------------------------------

\begin{figure}[!t]
\centering
\includegraphics[width=0.8\columnwidth]{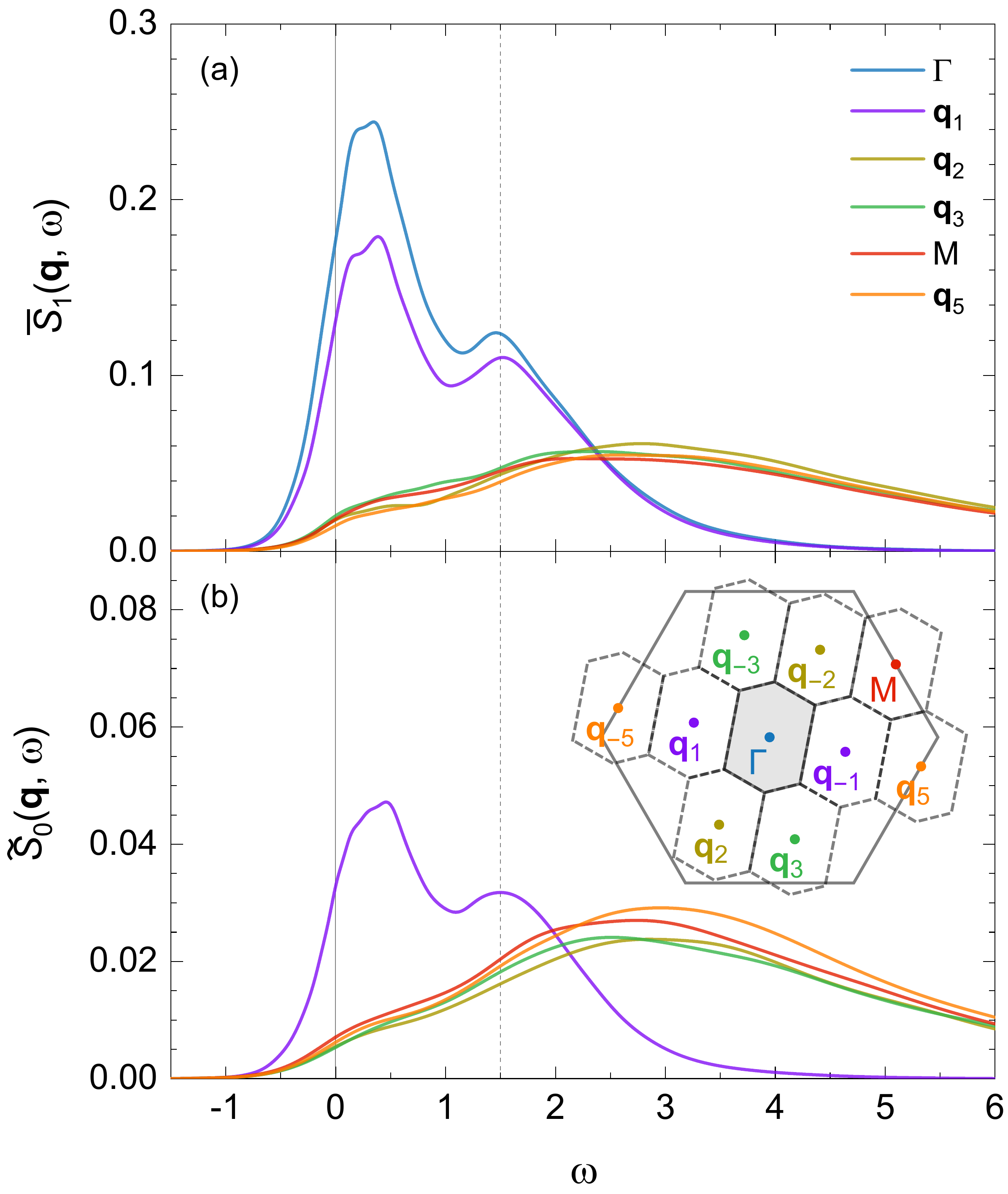}
\caption{(a) Average chiral DSF 
$\overline{S}_1({\bf q}, \omega)$ 
[\cref{eq_average_chiral_dsf}], and (b) chiral DSF $\widetilde{S}_0({\bf q}, \omega)$
at $T=0.2$ for all inequivalent numerical ${\bf q}$'s within the reduced BZ calculated on the $N=30$ lattice. Note the very different vertical scales of both panels. Inset in (b) shows the numerical ${\bf q}$-cells in the reduced BZ of the $N = 30$ lattice (\cref{fig1}).}
\label{fig3}
\end{figure}

In \cref{fig2} we present a comparison of chiral DSF's calculated at $T = 0.2$ using the FTLM on lattices with $N=24$, $27$ and $30$ sites (see \cref{fig1}). 
We choose two rather extreme cases of the $\Gamma$ and M points of the reduced BZ [see inset in \cref{fig3}(b)]. 
The largest dynamical response is at the $\Gamma$ point ($q=0$) with chirality $c=\pm 1$, which represents uniform fluctuations of the AFM order parameter for $\ang{120}$ ordered spins on each KL triangle. 
We see that these results are quite independent of lattice size $N$, 
which confirms that the spin correlation length is quite short even at this low temperature due to strong geometric frustration. 
The spectra are not featureless, as they exhibit two distinct frequency maxima, which seem quite robust.
These were already tentatively observed via the numerical linked cluster method \cite{sherman16} by assuming an \textit{ad hoc} Lorentzian line shape.
Our FTLM calculations, on the other hand, do not require any \textit{a priori} assumptions on the line shape.
The higher-energy maximum can be traced back to transitions within individual spin triangles, for which the energy gap between the $S=1/2$ GS and excited $S=3/2$ spin states 
%of an isolated spin triangle would be 
is 
$\omega=1.5$ (dashed line in \cref{fig2}). 

In \cref{fig3} we show the full chiral DSF $\widetilde{S}_c({\bf q}, \omega)$ at $T = 0.2$ 
for all inequivalent ${\bf q}$'s in the reduced BZ for both chirality branches $c=\pm 1$ and $c=0$,  calculated on the largest $N=30$ site lattice.
Since the $c=+1$ and $c=-1$ chiral DSF's are in general not equal at generic ${\bf q} \neq 0$ 
we plot in \cref{fig3}(a) the averaged chiral DSF
\begin{equation}
\overline{S}_1({\bf q}, \omega) = \frac{1}{2} \left[ \widetilde{S}_1({\bf q}, \omega) + \widetilde{S}_{-1}({\bf q}, \omega) \right] ,
\label{eq_average_chiral_dsf}
\end{equation}
while at $q = 0$ both chiralities $c = \pm 1$ match and we have $\overline{S}_1(q=0,\omega) = \widetilde{S}_{\pm 1}(q = 0, \omega)$. We see that chiral $c = \pm 1$ fluctuations indeed dominate the response (\cref{fig3}), with the largest intensity found at the $q=0$ ($\Gamma$) point and a slightly reduced intensity found at the smallest nonzero ${\bf q} = {\bf q}_1$.
Chiral DSF spectra at these low ${\bf q}$ show the characteristic double-maximum frequency dependence with maxima near $\omega \approx 0.3$  and $\omega \approx 1.5$ [\cref{fig3}(a)]. 
This structure is reproduced even in the considerably weaker $c = 0$ response at ${\bf q} = {\bf q}_1$ [\cref{fig3}(b)].
At larger ${\bf q}$, nearer the BZ boundary, all spectra are broad ($\delta \omega \gtrsim 3$), weak and featureless.
The observed ${\bf q}$ and $c$ dependence thus indicates that longer-ranged chiral $\ang{120}$ AFM correlations dominate the dynamical response of the KLHM at low frequencies, with a correlation length $\xi > 1$ extending further than a single spin triangle.

\begin{figure}[!b]
\centering
\includegraphics[width=0.8\columnwidth]{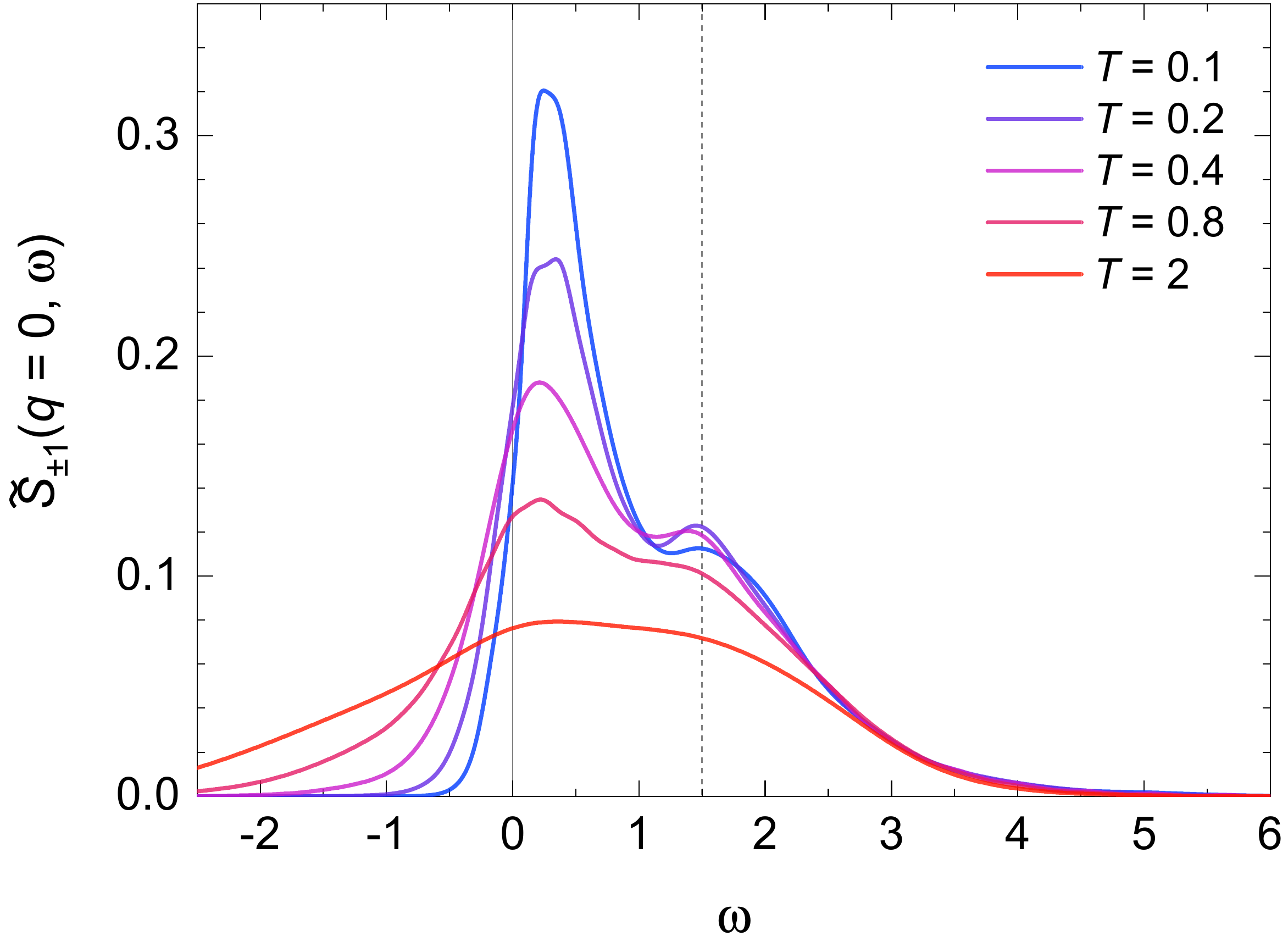}
\caption{The chiral DSF $\widetilde{S}_{\pm 1} (q=0, \omega)$ at different $T = 0.1$--$2.0$ on the $N=30$ lattice.}
\label{fig4}
\end{figure}

In \cref{fig4} we present the temperature evolution of the dominant $q = 0$, $c = \pm 1$ chiral DSF. It is evident that the double-maximum frequency structure is not just a low-$T$ feature 
as it persists to temperatures as large as $T \sim 1$.
Partly, the low-energy peak at $\omega \sim 0.3$ is simply a consequence of the detailed balance relation for DSF's, $S(-\omega) = \exp(-\omega/T) S(\omega)$,
which implies $\dd S / \dd \omega|_{\omega = 0} = S(0) / (2 T) > 0$ and thus always leads to a maximum at $\omega > 0$.
On the other hand, at the lowest $T = 0.1 \sim T_\mathrm{fs} $ we find a further reduced $\omega = 0$ response, which could indicate a finite spin triplet gap $\Delta_t > 0$, at least on our finite-sized $N = 30$ lattice.

Finally, we note that the $q = 0$, $c = 0$ chiral DSF had to be explicitly excluded from our FTLM calculations since it is singular in finite systems,  $\widetilde{S}_0(q = 0, \omega) \propto \delta(\omega)$, 
due to the conservation of $S^z_\mathrm{tot}$.
Nevertheless, in the macroscopic limit $N \to \infty$ at $T > 0$ this singular DSF should evolve into the $q \approx 0$ spin diffusion peak with a spectral width that is expected to scale as $\delta \omega \propto D_\mathrm{diff} q^2$ where $D_\mathrm{diff}(T)$ is the temperature-dependent spin diffusion constant \cite{sokol93}.
We discuss the experimental relevance of this contribution in more detail in \cref{sec5b}.

\subsection{Equal-time correlations and static response}\label{sec3b}

\begin{figure}[!t]
\centering
\includegraphics[width=\columnwidth]{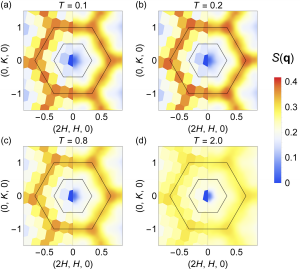}
\caption{The equal-time spin correlation function $S({\bf q})$ in the extended BZ (large hexagon) at different $T=0.1$--$2.0$ on the $N = 30$ lattice. The reduced BZ is shown by the smaller hexagon. The left halves of the panels show raw FTLM results on ${\bf q}$-cells shown in the inset in \cref{fig3}(b), while the right halves are rotationally symmetrized (to recover the threefold symmetry of the KL) and smoothed via interpolation.}
\label{fig5}
\end{figure}

For comparison with experimental INS data as well as with previous theoretical calculations, it is informative to also look at the standard INS DSF $S({\bf q}, \omega)$ [\cref{eq_dsfs}] in the extended BZ, 
which is calculated from the chiral DSF $\widetilde{S}_c({\bf q},\omega)$ via \cref{eq_dsf_from_chiral_dsf}.
Firstly, we consider the ${\bf q}$-dependent equal-time spin correlation function $S({\bf q})$  [\cref{eqs_quantities}] over a broad range of temperatures $T = 0.1$--$2.0$ on the $N=30$ lattice (\cref{fig5}).
Consistent with several previous numerical studies of this quantity \cite{iqbal13,shimokawa16,sherman18}, $S({\bf q})$ has a pronounced but spread-out region of high intensity around the whole extended BZ boundary that remains visible even at very high $T \sim 2$. 
This can be understood by considering the ${\bf q}$-dependence of the chiral weighing factor $|\xi_c({\bf q})|^2$ in \cref{eq_dsf_from_chiral_dsf} that suppresses the contribution of the dominant chiral $c = \pm 1$ fluctuations to the standard DSF $S({\bf q}, \omega)$ near the $\Gamma$ point of the extended BZ, but not at the extended BZ boundary.
Weak global maxima of $S({\bf q})$ appear for $T>0.1$ at corner K points of the extended BZ (note that our $N=30$ 
lattice does not contain this point),  qualitatively consistent with previous studies, 
but appreciable intensity can also be found at the M points (corresponding to periodic images of the $\Gamma$ point of the reduced BZ).

\begin{figure}[!t]
\centering
\includegraphics[width=\columnwidth]{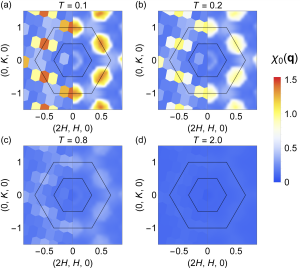}
\caption{The d.c. spin susceptibility $\chi_0({\bf q})$ in the extended BZ (large hexagon) at different $T=0.1$--$2.0$ on the $N = 30$ lattice. The reduced BZ is shown by the smaller hexagon. The left halves of the panels show raw FTLM results on ${\bf q}$-cells shown in the inset in \cref{fig3}(b), while the right halves are rotationally symmetrized as in \cref{fig5}.}
\label{fig6}
\end{figure}

A complementary quantity, which is more sensitive to low-energy fluctuations as is obvious from \cref{eqs_quantities}, is the ${\bf q}$-dependent d.c.~susceptibility $\chi_0({\bf q})$, which we present in the extended BZ over a broad range of temperatures in \cref{fig6}.
A striking difference to $S({\bf q})$ (\cref{fig5}) is a very pronounced maximum of $\chi_0({\bf q})$ at the M point of the extended BZ, which is directly
related to the dominant low-energy $q = 0$, $c = \pm 1$ chiral fluctuations seen in the chiral DSF (\cref{fig3}). 
This maximum is much more sensitive to temperature than the maximum in $S({\bf q})$ and disappears for $T>1$,
consistent with the broadening of the chiral response visible in \cref{fig4}. 
It should be stressed that the same maximum is directly related to the one observed by low-energy INS in herbertsmithite \cite{han12}, as will be discussed in more detail in \cref{sec5a}.

\subsection{Local spin fluctuations}\label{sec3c}

LSF can be expressed from the chiral DSF as
\begin{equation}
S_L^{\alpha\alpha}(\omega) = \int_{-\infty}^\infty \frac{\dd t}{2\pi}~ e^{i \omega t} \braket{S_i^\alpha(t) S_i^\alpha(0)} = \frac{1}{N}\sum_{c {\bf q}} \widetilde{S}_c^{\alpha\alpha}({\bf  q},\omega) ,
\label{eq_sl}
\end{equation}
and are likewise isotropic in the $D=0$ KLHM, i.e. $S_L^{\alpha\alpha}(\omega) = S_L(\omega)$.
Their value at $\omega \approx 0$ is experimentally highly relevant, as it is directly proportional to the experimental NMR spin-lattice relaxation rate $1/T_1$, 
provided that hyperfine form factors do not play an essential role, as discussed in detail in \cref{sec5b}. 
As mentioned previously, we omit the singular $q=0$, $c=0$ spin diffusion contribution, discussed further in \cref{sec5b}. 

\begin{figure}[!t]
\centering
\includegraphics[width=0.8\columnwidth]{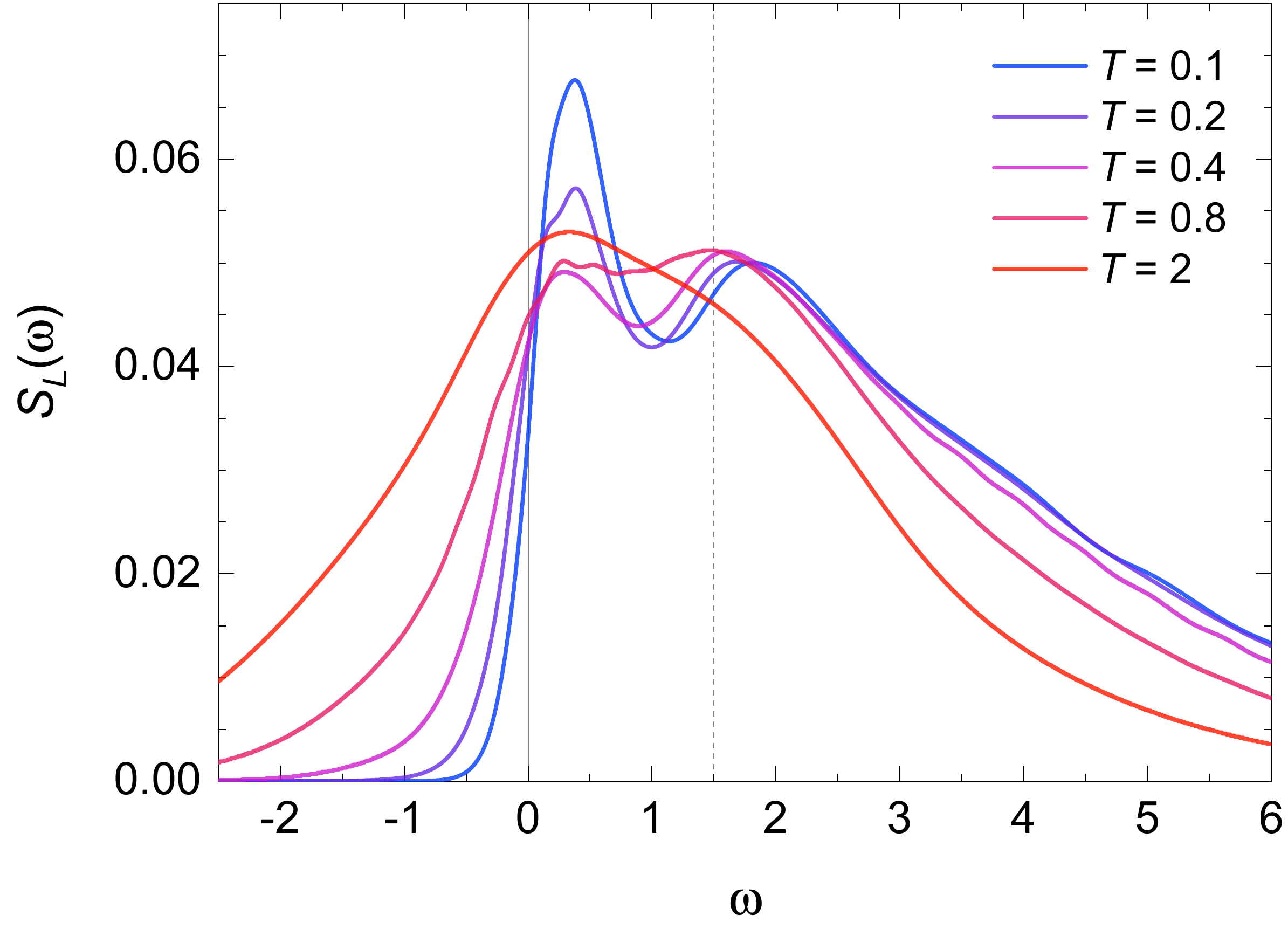}
\caption{The LSF $S_L(\omega)$ at different $T = 0.1$--$2.0$ on the $N=30$ lattice.}
\label{fig7}
\end{figure}

In \cref{fig7} we show the temperature evolution of the LSF over a broad range of $T = 0.1$--$2.0$ on the $N=30$ lattice. 
Apart from a pronounced low-energy peak arising from the dominant $q = 0$, $c = \pm 1$ chiral fluctuations at $T < 0.2$ the LSF are quite temperature independent for $T>0.2$, even at the relevant $\omega \approx 0$ energy scale of NMR experiments.  At $T < 0.2$, a drop of $S_L(\omega = 0)$ is observed, 
which is again a signature of a finite spin triplet gap $\Delta_t >0$, at least on finite-sized lattices \cite{schnack18,lauchli19,prelovsek20}.

It is instructive to compare the calculated LSF $S_{L}(\omega = 0)$ to Moriya's Gaussian approximation \cite{moriya56} frequently used at high $T \gg 1$,
but also extended to lower $T$ via higher-order corrections in the case of the $D = 0$ KLHM \cite{sherman16}. 
In a uniform Heisenberg spin-$1/2$ model the LSF frequency moments 
$\mu_k = \int d \omega~ \omega^k S_L(\omega) $,
are exactly known at $T \to \infty$, with the LSF sum rule $\mu_0 = 1/4$ and $\mu_2 = z/8$, 
where $z=4$ is the number of nearest-neighbors in the KL.
These yield the expected $\omega = 0$ value of the KLHM LSF under Gaussian line shape approximation
\begin{equation}
S_{L}^\mathrm{Moriya}(0) = \frac{\mu_0}{\sqrt{8 \pi \mu_2}} \approx 0.071 ,
\label{eq_moriya_SL}
\end{equation}
which is reasonably close to the actual KLHM value $S_{L}(0) \approx 0.055$ at $T = 2$ calculated with the FTLM.
We note, though, that the frequency-dependent LSF $S_{L}(\omega)$ are not, in fact, Gaussian in shape, as is obvious from \cref{fig7}, 
and $T=2$ is not yet ${\gg}1$.

\section{Dzyaloshinskii-Moriya interactions}\label{sec4}

In this section we consider an extension of the KLHM with out-of-plane DM interactions $0 \leq D \leq 0.25$
[\cref{eq_hm}] (note that the dynamical response is not sensitive to the sign of $D$) \cite{rigol072}, which are relevant in many KL materials \cite{zorko08,el10,zorko13,zorko192,arh20}. 
The out-of-plane $D$ leads to a uniaxially anisotropic chiral DSF $\widetilde{S}_c^{\alpha\alpha}({\bf q},\omega)$ with equal $\alpha = x$ and $\alpha = y$ components that differ from the $\alpha = z$ component, which has to be calculated separately.
The same also holds for all derived quantities, including the standard DSF $S^{\alpha\alpha}({\bf q},\omega)$ [\cref{eq_dsfs}] and quantities in \cref{eqs_quantities}.
We note that chiral spin operators $\widetilde{S}^\alpha_{c{\bf q}}$ with $\alpha = x,y$ are off-diagonal in the $S^z_\mathrm{tot}$ basis, which 
substantially increases the overall computational complexity and requirements of FTLM compared to the $\alpha = z$ case, 
where they are diagonal in the subspace. 
In particular, the employed reduced summation over $S^z_\mathrm{tot}$ subspaces (having lesser effect on diagonal correlations) appears to influence more the calculation of off-diagonal $\alpha = x,y$ correlations. To reduce differences we normalize $\alpha = x,y$ results by a scaling factor of $1.15$ to reproduce the $\alpha = z$ 
sum rules at $D=0$.

\subsection{Dynamical spin structure factor}\label{sec4a}

\begin{figure}[!t]
\centering
\includegraphics[width=0.8\columnwidth]{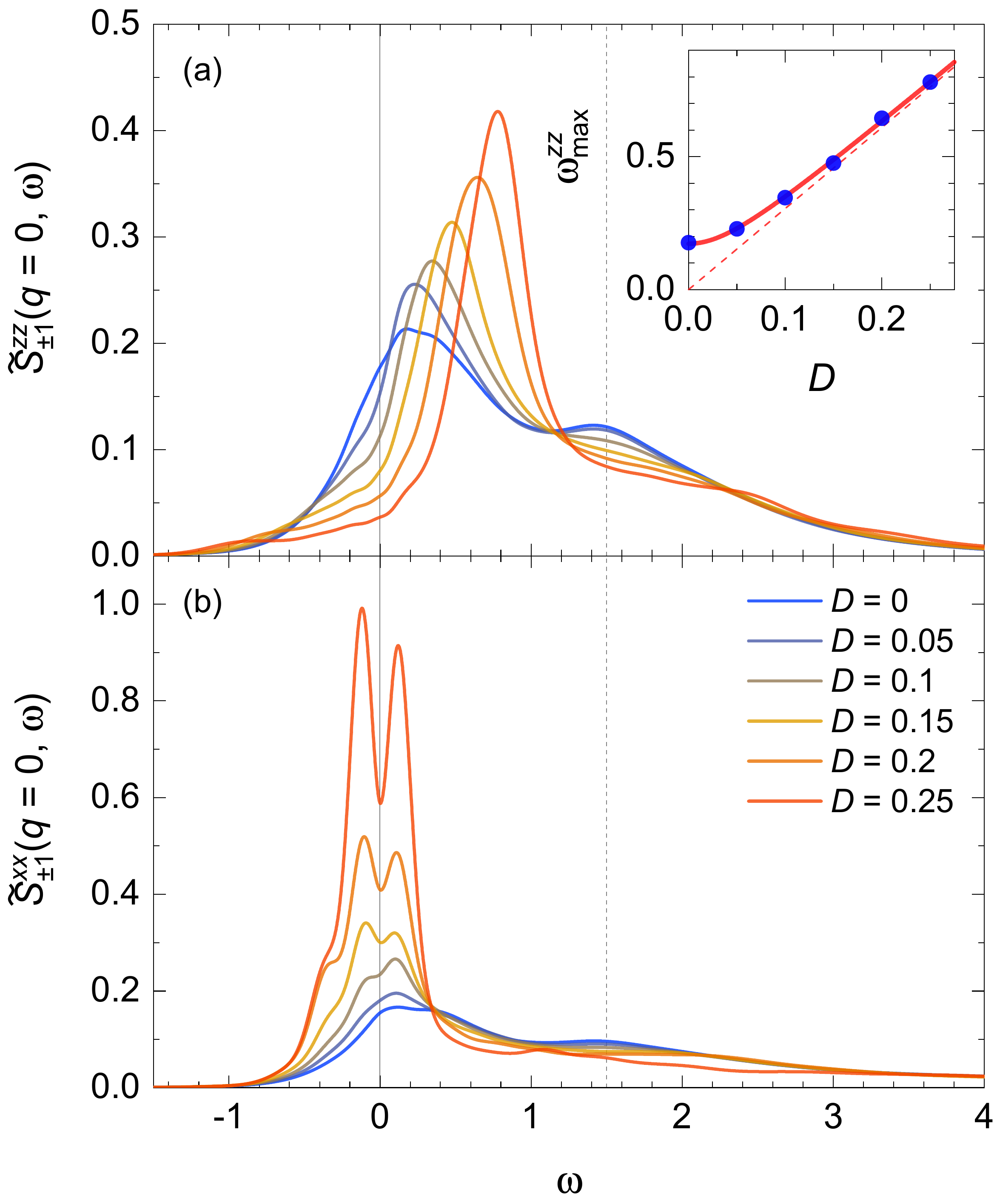}
\caption{Chiral DSF's (a) $\widetilde{S}^{zz}_{\pm 1}(q=0, \omega)$ and (b) $\widetilde{S}^{xx}_{\pm 1}(q=0, \omega)$ at a fixed $T = 0.3$ and different $D = 0$--$0.25$ on the $N = 30$ lattice.
Inset in (a) shows the frequency of the lower-energy maximum of $\widetilde{S}^{zz}_{\pm 1}(q=0, \omega)$ (symbols) with curves serving as guides to the eye.}
\label{fig8}
\label[pluralfigure]{figs8}
\end{figure}

It is known that at low temperatures KLHM systems can be significantly affected by the presence of additional DM interactions, with a quantum phase transition from a SL GS to a $\ang{120}$ AFM LRO GS with nonzero vector spin chirality when $D > D_c \approx 0.1$ \cite{cepas08,elhajal02}. 
This mainly corresponds to a gradual softening of the dominant $q = 0$, $c = \pm 1$ chiral fluctuations as $D$ increases towards the quantum critical point $D_c$, beyond which these emerge as in-plane chiral $\ang{120}$ AFM LRO.

In \cref{fig8} we present the dominant $q = 0$, $c = \pm 1$ chiral DSF at a temperature $T = 0.3$ high enough to avoid 
longer-ranged AFM correlations leading to strong finite-size effects in our FTLM calculations.
A finite $D > 0$ substantially decreases the $\alpha = z$ component of the chiral DSF at low $\omega$, consistent with an increase of the effective out-of-plane spin triplet gap $\Delta^z_t$. 
At the same time, the $\alpha = z$ spectra become sharper (more coherent) for $D > 0.1$, i.e.~beyond the quantum critical point, with the energy of the spectral peak scaling nearly linearly as 
$\omega_\mathrm{max} \approx 3.0 D$ [see inset in \cref{fig8}(a)].
This is consistent with the linear scaling of the lower specific heat peak $T_\mathrm{max} \approx 0.91 D$ 
found via the FTLM in Ref.~\cite{arh20}, 
below which the spin correlation length $\xi$ increases substantially.
The $\alpha = x,y$ components of the chiral DSF show the latter effect quite clearly [\cref{fig8}(b)] 
as low-energy oscillations due to finite-size magnon-like excitations 
become visible at $D \gtrsim 0.1$ and increase in prominence as $D$ increases further. 
This indicates a considerable increase in the spin correlation length $\xi > 1$ already at $T \gtrsim T_\mathrm{max}$ with increasing $D > D_c$.  Concomitantly, there is a substantial increase of low-$\omega$ intensity in 
$\alpha = x,y$ components of the chiral DSF, 
in contrast to a decrease in the $\alpha = z$ component, 
consistent with a softening of chiral fluctuations above a $\ang{120}$ AFM GS with in-plane LRO spins \cite{cepas08,elhajal02,zorko08,zorko13,zorko192,arh20}.

\subsection{Equal-time correlations and local spin fluctuations}\label{sec4b}

\begin{figure}[!t]
\centering
\includegraphics[width=0.8\columnwidth]{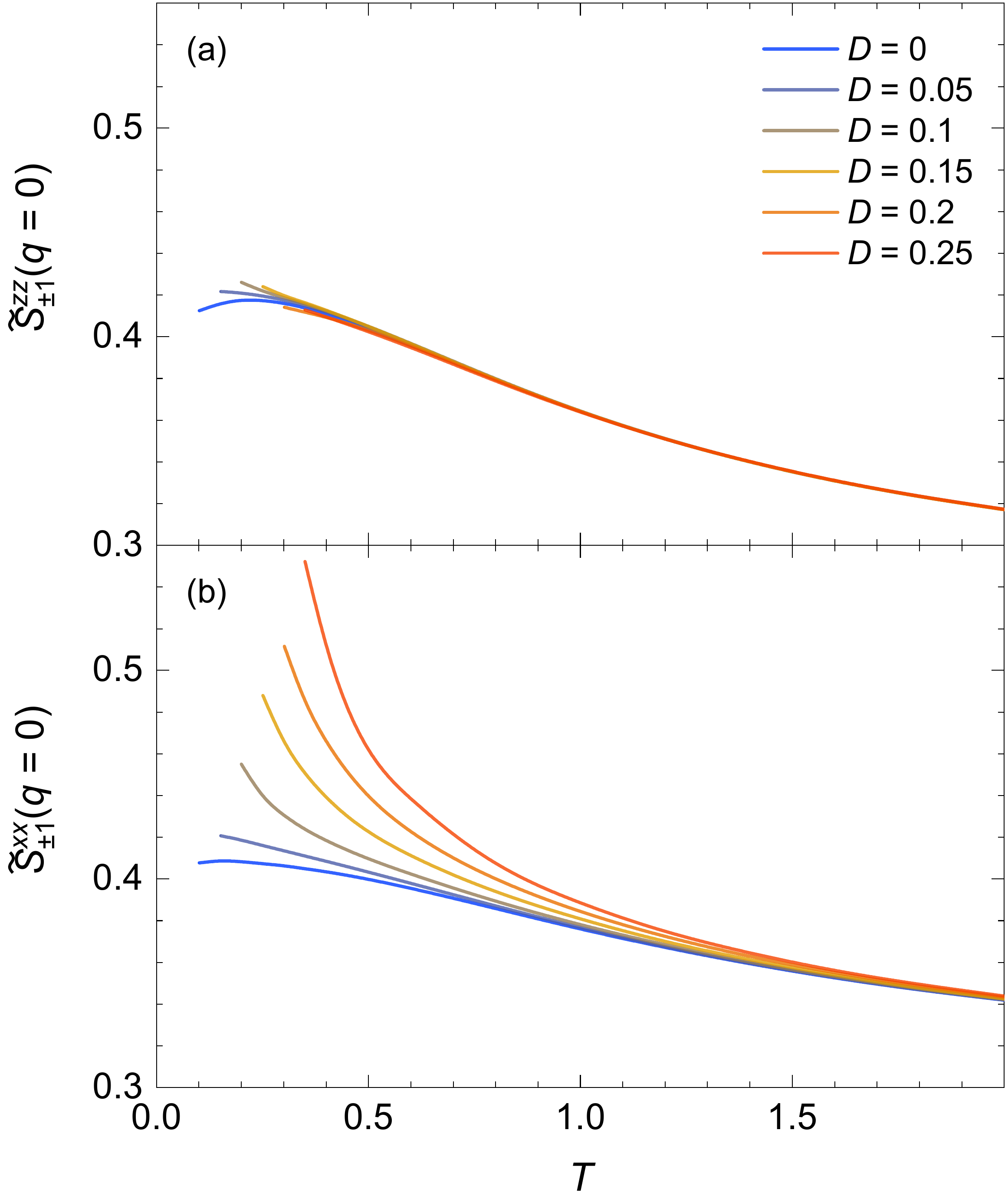}
\caption{The temperature dependence of chiral equal-time spin correlation functions (a) $\widetilde{S}^{zz}_{\pm 1}(q=0)$ and (b) $\widetilde{S}^{xx}_{\pm 1}(q=0)$ for different $D = 0$--$0.25$ on the $N = 27$ lattice. }
\label{fig9}
\label[pluralfigure]{figs9}
\end{figure}

Similar conclusions can be drawn from the temperature dependence of the chiral equal-time correlation function 
$\widetilde{S}^{\alpha\alpha}_c({\bf q})$, which is 
defined by replacing the standard DSF $S^{\alpha\alpha}({\bf q},\omega)$ in \cref{eqs_quantities} by the chiral DSF $\widetilde{S}^{\alpha\alpha}_c({\bf q},\omega)$.
We focus on the dominant $q = 0$, $c = \pm 1$ correlations, which are shown in \cref{fig9}. 
We see that they are weakly $T$-dependent over the whole $T> T_\mathrm{fs} $ range  when $D < D_c$.  The behavior changes qualitatively for $D > D_c$. While the $\alpha = z$ component remains 
relatively unaffected [\cref{fig9}(a)], 
the $\alpha = x,y$ components show a strong increase below $T \lesssim 2D$ [\cref{fig9}(b)] 
consistent with the gradual onset of longer-range correlations around $T \sim T_\mathrm{max}$ \cite{arh20} and ultimate chiral AFM LRO at $T = 0$.

\begin{figure*}[!t]
\centering
\includegraphics[width=0.8\linewidth]{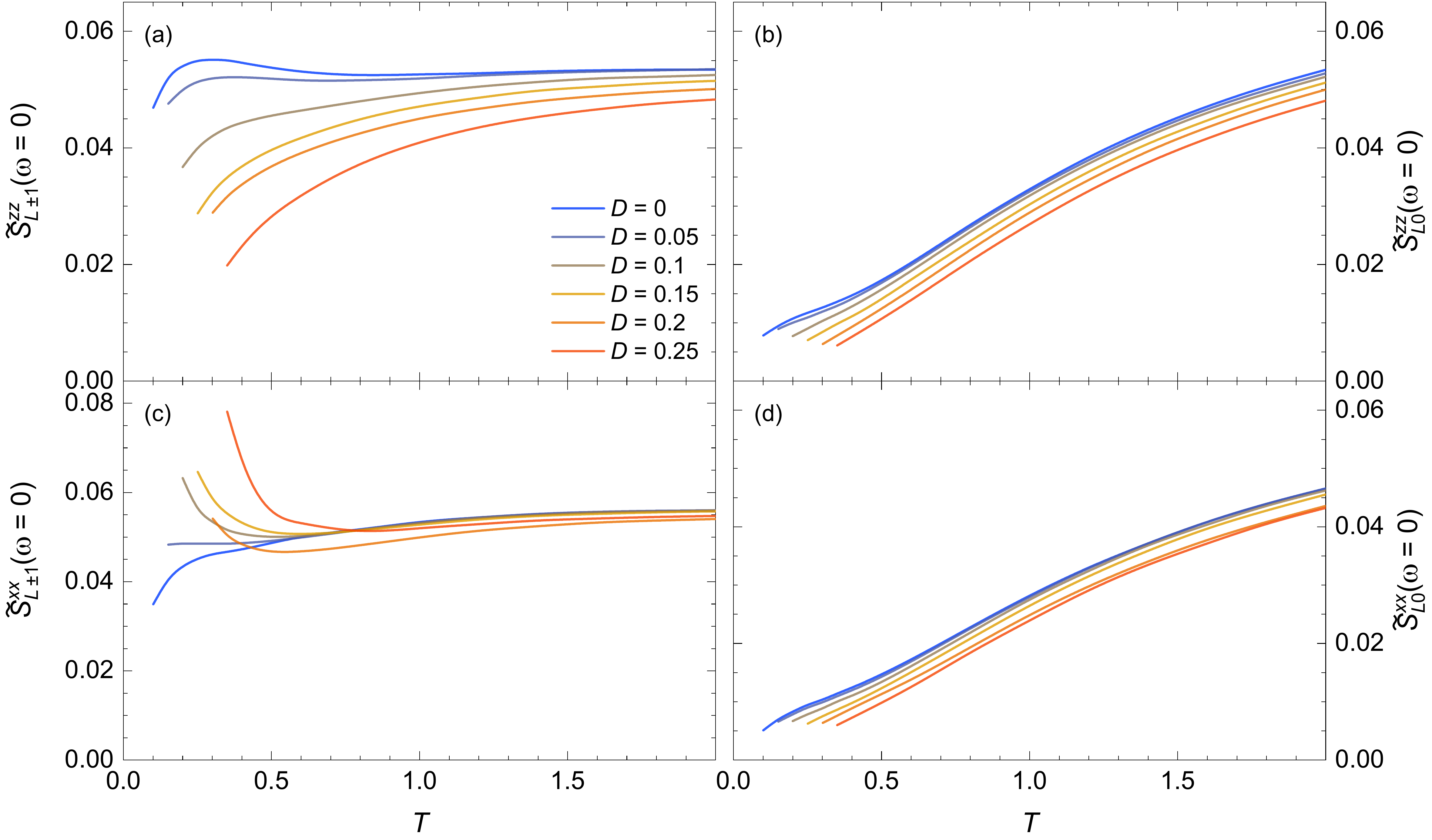}
\caption{The temperature dependence of d.c. chiral LSF (a) $\widetilde{S}^{zz}_{L {\pm}1}(\omega=0)$, (b) $\widetilde{S}^{zz}_{L0}(\omega=0)$, (c) $\widetilde{S}^{xx}_{L {\pm}1}(\omega=0)$, and (d) $\widetilde{S}^{xx}_{L0}(\omega=0)$ for different $D=0$--$0.25$ on the $N = 30$ lattice.}
\label{fig10}
\label[pluralfigure]{figs10}
\end{figure*}

Finally, we consider the temperature dependence of the $\omega = 0$ LSF, which are directly relevant for NMR spin-lattice relaxation rate ($1/T_1$) experiments that we discuss in \cref{sec5b}. 
Here we find it useful to separately consider the individual chiral LSF contributions
\begin{equation}
\widetilde{S}^{\alpha\alpha}_{Lc}(\omega) = \frac{3}{N}\sum_{\bf q} \widetilde{S}^{\alpha\alpha}_c({\bf  q},\omega) ,
\label{eq_sl_chiral}
\end{equation}
to the full LSF $S_L^{\alpha\alpha}(\omega) = (1/3) \sum_c \widetilde{S}^{\alpha\alpha}_{Lc}(\omega)$.
Note that the $c=+1$ and $c=-1$ chiral LSF are equal.  

In \cref{fig10} we present the calculated temperature dependence of the $\omega = 0$ chiral LSF for a range of $D = 0$--$0.25$ on the $N = 30$ lattice. 
We find that the $c = \pm 1$ chiral LSF are highly sensitive to $D$, 
especially at $T \lesssim 2D$ 
where the $\alpha = z$ component is suppressed [\cref{fig10}(a)] due to a shift of spectral intensity to higher $\omega \sim \omega_\mathrm{max}$ [\cref{fig8}(a)],
while the $\alpha = x,y$ components are strongly enhanced [\cref{fig10}(c)] due to the gradual onset of longer-range correlations at $T \sim T_\mathrm{max}$ [\cref{fig8}(b)]. 
On the other hand, components of the $c = 0$ chiral LSF are nearly equal and mostly insensitive to $D$, showing just a steady increase with increasing temperature due to increasingly incoherent spin dynamics at $T \gtrsim 1$ [\cref{fig10}(b,d)].

\section{Comparison with experiment}\label{sec5}

In this section we reinstate $J \neq 1$ and SI units. 
%Compared to previous sections, this implies $\omega \rightarrow \hbar\omega/J$, $T \rightarrow k_B T/J$ and $D \rightarrow D/J$. 

\subsection{Inelastic neutron scattering}\label{sec5a}

INS is a very powerful experimental technique as it directly probes the magnetic DSF $S^{\alpha\alpha}({\bf q},\omega)$, 
with typical interaction energies in KLHM materials $J \sim k_B (60$--$\SI{230}{K}) = 5$--$\SI{20}{meV}$ in a convenient energy range for this technique. 
Unfortunately, most KL materials are not yet available in single-crystal form, 
therefore the intrinsic DSF anisotropy and ${\bf q}$-dependence is often averaged out in experiment. 
To avoid this issue we concentrate on INS results on single-crystal herbertsmithite \cite{han12}, 
a material that remains in a SL state down to the lowest measured $k_B T \approx 10^{-4}J$. 
As the experimentally determined $D=(0.04$--$0.08)J$ \cite{zorko08,el10} plays only a modest role against a much stronger $J/k_B \approx \SI{190}{K}$ ($\SI{16.4}{meV}$) in equal-time properties relevant for INS (\cref{fig9}), 
we compare INS results with model calculations for $D = 0$. 
Moreover, low-$\omega$ results may be strongly influenced by structural and chemical disorder, especially at low $k_B T \ll J$. 
We therefore restrict ourselves to INS energies $\hbar\omega > \SI{1}{meV}$, 
above the energy scale of impurity contributions, 
which mostly contribute to a low-$\omega$ quasielastic INS peak \cite{han12,han16}. 

\begin{figure}[!b]
\centering
\includegraphics[trim = 0mm 0mm 0mm 0mm, clip, width=0.8\columnwidth]{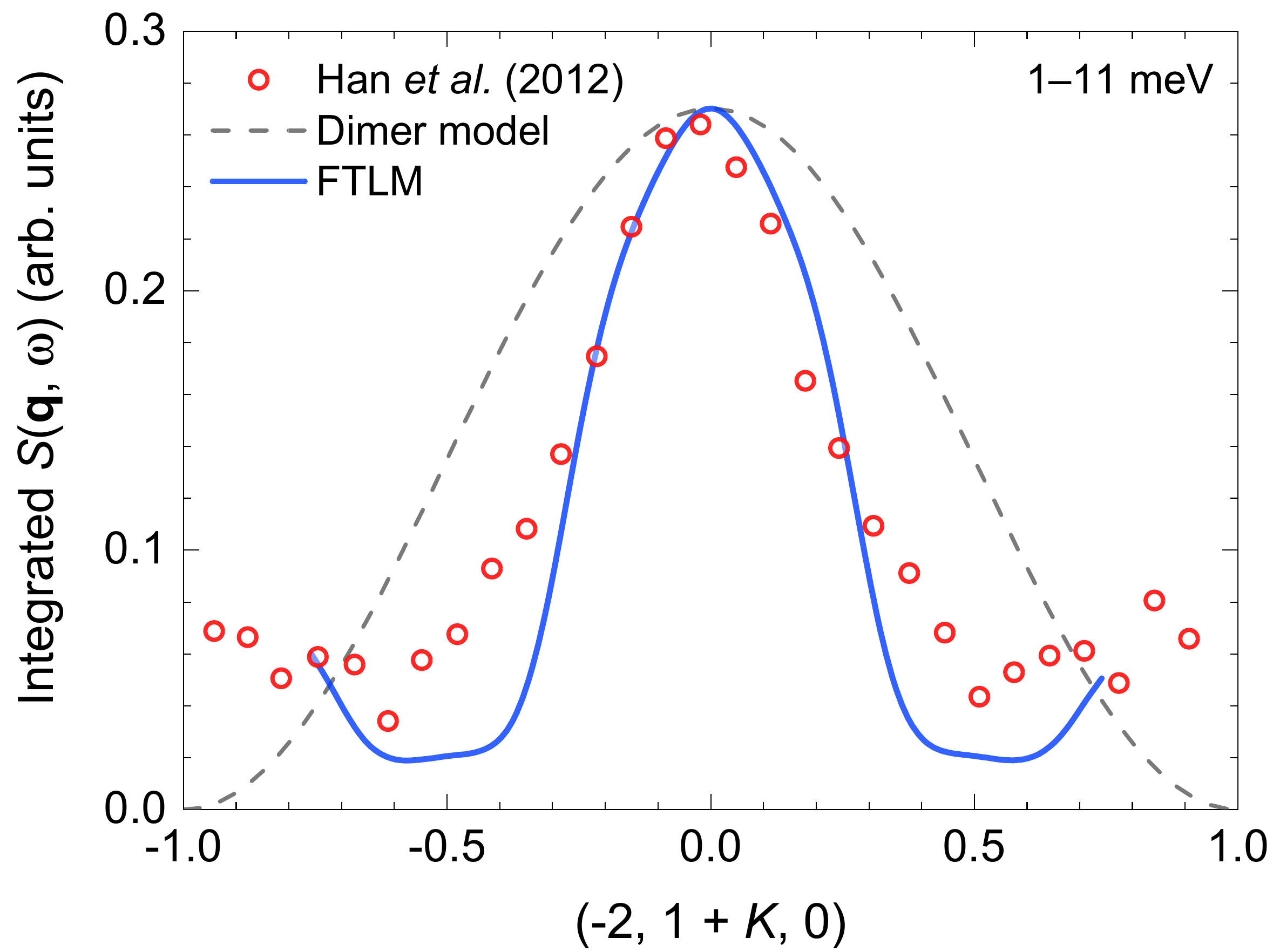}
\caption{Low-$T$ herbertsmithite INS measurements of the magnetic DSF $S({\bf q}, \omega)$ (symbols) integrated over $\SI{1}{meV} < \hbar\omega < \SI{11}{meV}$ along the $(-2, 1 + K, 0)$ cut in ${\bf q}$-space from Ref.~\cite{han12}. 
The presented magnetic DSF was obtained from raw experimental data by dividing INS intensities by the free-\ch{Cu^{2+}} magnetic form factor $|F({\bf q})|^2$ \cite{han12,shirane06}. 
The experimental magnetic DSF agrees well with ${\bf q}$-interpolated FTLM calculations at $T = 0.1$ on the $N = 30$ lattice (blue line), but significantly worse with a toy model of independent singlet dimers \cite{han12} (dashed line).}
\label{fig11}
\end{figure}

Firstly, we note that experimental frequency-dependent INS spectra show a broad maximum at $\hbar\omega \sim \SI{6}{meV}$ \cite{han12}, 
which is consistent with the calculated low-energy peak at $\hbar\omega \approx 0.3 J = \SI{5}{meV}$ (\cref{fig3,fig4}).
Secondly, our calculations also nicely reproduce the ${\bf q}$-dependence of the DSF integrated over a broad frequency window $\SI{1}{meV} < \hbar\omega < \SI{11}{meV}$ (i.e.~$0.06 J < \hbar\omega < 0.67 J$) along the $(-2, 1 + K, 0)$ cut in ${\bf q}$-space, 
which shows the most pronounced variation (\cref{fig11}). 
The position of the experimental INS peak at $(-2, 1, 0)$ corresponds to the M point of the extended BZ and is well accounted for by our FTLM results. 
Having separated chiral contributions at different wave vectors, we can attribute this peak to the dominant low-energy $q = 0$, $c = \pm 1$ chiral fluctuations [\cref{fig3,eq_dsf_from_chiral_dsf}]. 
The position of the peak is also consistent with expectations from the ${\bf q}$-dependent d.c.~susceptibility $\chi_0({\bf q})$, 
which is also sensitive mainly to low-energy fluctuations, 
and which also has a very pronounced peak at the same wavevector (\cref{fig6} and discussion in \cref{sec3b}).
Finally, we stress that not only the position but also the width of the experimental INS peak is well reproduced by model calculations (\cref{fig11}), 
and is considerably smaller than the width predicted by a simple independent singlet dimer model \cite{han12}.
This indicates that the chiral AFM fluctuations in the KLHM have a nontrivial low-$T$ correlation length $\xi > 1$ that extends beyond nearest KL neighbors.

\subsection{NMR spin-lattice relaxation rate}\label{sec5b}

\subsubsection{Theory}

NMR spin relaxation experiments probe low-energy electron spin fluctuations via the hyperfine coupling between nuclear and electron spins. 
In a crystal, the spin-lattice relaxation rate of a given nucleus is given by \cite{horvatic92,horvatic01}
\begin{equation}
\frac{1}{T_1} = \frac{\gamma_n^2}{2} \int_{-\infty}^\infty \dd t~ e^{i \omega_0 t} \sum_{i j \alpha \beta} (\delta_{\alpha\beta} - \hat{B}_\alpha \hat{B}_\beta) \braket{{\delta b}_i^\alpha(t) {\delta b}_j^\beta(0)} ,
\label{eq_t1}
\end{equation}
where $\gamma_n$ is the nuclear gyromagnetic ratio, $\omega_0 = \gamma_n B \ll J/ \hbar$ is the nuclear Larmor angular frequency in an external field ${\bf B}$, $\hat{\bf B} = {\bf B}/|{\bf B}|$ is a unit vector pointing along ${\bf B}$,
and ${\delta b}_i^\alpha = b_i^\alpha - \braket{b_i^\alpha} = -\sum_\mu A_i^{\alpha\mu} S_i^\mu$ is the effective fluctuating local field at the position of the nucleus due to hyperfine coupling with the electron spin $S_i^\mu$ via the specific hyperfine coupling tensor $A_i^{\alpha\mu}$.
Defining the chiral hyperfine coupling tensor in ${\bf q}$-space in analogy with [\cref{eq_sqc}] as
\begin{equation}
\widetilde{A}_{c{\bf q}}^{\alpha\mu} = \frac{1}{\sqrt{N}} \sum_n e^{i {\bf q} \cdot \widetilde{\bf R}_n} \left[ A_{(n,0)}^{\alpha\mu} + \zeta^c A_{(n,1)}^{\alpha\mu} + \zeta^{-c} A_{(n,2)}^{\alpha\mu} \right] , 
\label{eq_aqc}
\end{equation}
we can further succinctly express the NMR spin-lattice relaxation rate in terms of the chiral DSF $\widetilde{S}_c^{\alpha\beta}({\bf q},\omega)$ [\cref{eq_chiral_dsf}] as
\begin{equation}
\frac{1}{T_1} = \pi \gamma_n^2 \sum_{c{\bf q}} \tr \left\{ \widetilde{\underbar{A}}_{c{\bf q}}^\dagger \cdot \underbar{P}_\perp \cdot \widetilde{\underbar{A}}_{c{\bf q}} \cdot \widetilde{\underbar{S}}_c({\bf q},\omega_0) \right\} ,
\label{eq_t1_chiral_dsf_tensorial}
\end{equation}
where the tensor $\underbar{P}_\perp = \underbar{I} - \hat{\bf B} \otimes \hat{\bf B}$ projects onto a plane orthogonal to ${\bf B}$, while $\widetilde{\underbar{A}}_{c{\bf q}}$ and $\widetilde{\underbar{S}}_c({\bf q},\omega_0)$ are $3 \times 3$ tensors with components $\widetilde{A}_{c{\bf q}}^{\alpha\mu}$ and $\widetilde{S}_c^{\alpha\beta}({\bf q},\omega_0)$, respectively.

\begin{ruledtabular}
\begin{table}[!b]
\caption{NMR chiral form factors $f_c$ in \cref{eqs_t1_chiral_fs} for coupling to different numbers of spins $z_1$ in a single KL triangle and examples of relevant nuclei in herbertsmithite, \ycuFormula and other compounds [see inset in \cref{fig12}(a)]. Note that $\sum_c f_c = 3 z_1$.}
\begin{center}
\renewcommand{\arraystretch}{1.2}
\begin{tabular}{@{} r l l c c @{}}
$z_1$ & Nuclear position & Nucleus & Non-chiral $f_0$ & Chiral $f_{\pm 1}$ \\
\midrule
1 & Magnetic ion (on-site)  & \ch{^{63{,}65}Cu}       & 1 & 1 \\%& On-site\\
2 & Exchange bond (NN)      & \ch{^{17}O}, \ch{^{1}H} & 4 & 1 \\%& On-site \& pair\\
3 & Center of spin triangle & \ch{^{35}Cl}            & 9 & 0 \\%& On-site \& pair\\
\end{tabular}
\end{center}
\label{table_NMR_fc}
\end{table}
\end{ruledtabular}

In the simplest, yet experimentally highly relevant, case of a nucleus coupled to $z_1$ spins of the KL triangle 
with hyperfine eigenaxes along the crystallographic axes 
[i.e.~for $A_{(n,k)}^{\alpha\mu} = A_\alpha \delta_{\alpha\mu} \delta_{n,n_0} \delta_{k \leq z_1}$], 
this further simplifies to 
an expression involving only the chiral LSF $\widetilde{S}_{L c}^{\alpha\alpha}(\omega_0)$ [\cref{eq_sl_chiral,fig10}]
\begin{equation}
\begin{split}
\frac{1}{T_1} &= \pi \gamma_n^2 \sum_\alpha A_\alpha^2 ( 1 - \hat{B}_\alpha^2 ) \frac{1}{\widetilde{T}_1^{\alpha\alpha}} ,\\
\frac{1}{\widetilde{T}_1^{\alpha\alpha}} &= \frac{1}{3} \sum_{c=-1}^1 f_c \widetilde{S}_{Lc}^{\alpha\alpha}(\omega_0) ,
\end{split}
\label[pluralequation]{eqs_t1_chiral_fs}
\end{equation}
where $1/\widetilde{T}_1^{\alpha\alpha}$ are directional contributions to the spin-lattice relaxation rate $1/T_1$ 
that depend on the number of spins $z_1$ the nucleus is coupled to via the chiral form factors $f_c$ summarized in \cref{table_NMR_fc}. 

\begin{figure*}[!t]
\centering
\includegraphics[width=0.8\linewidth]{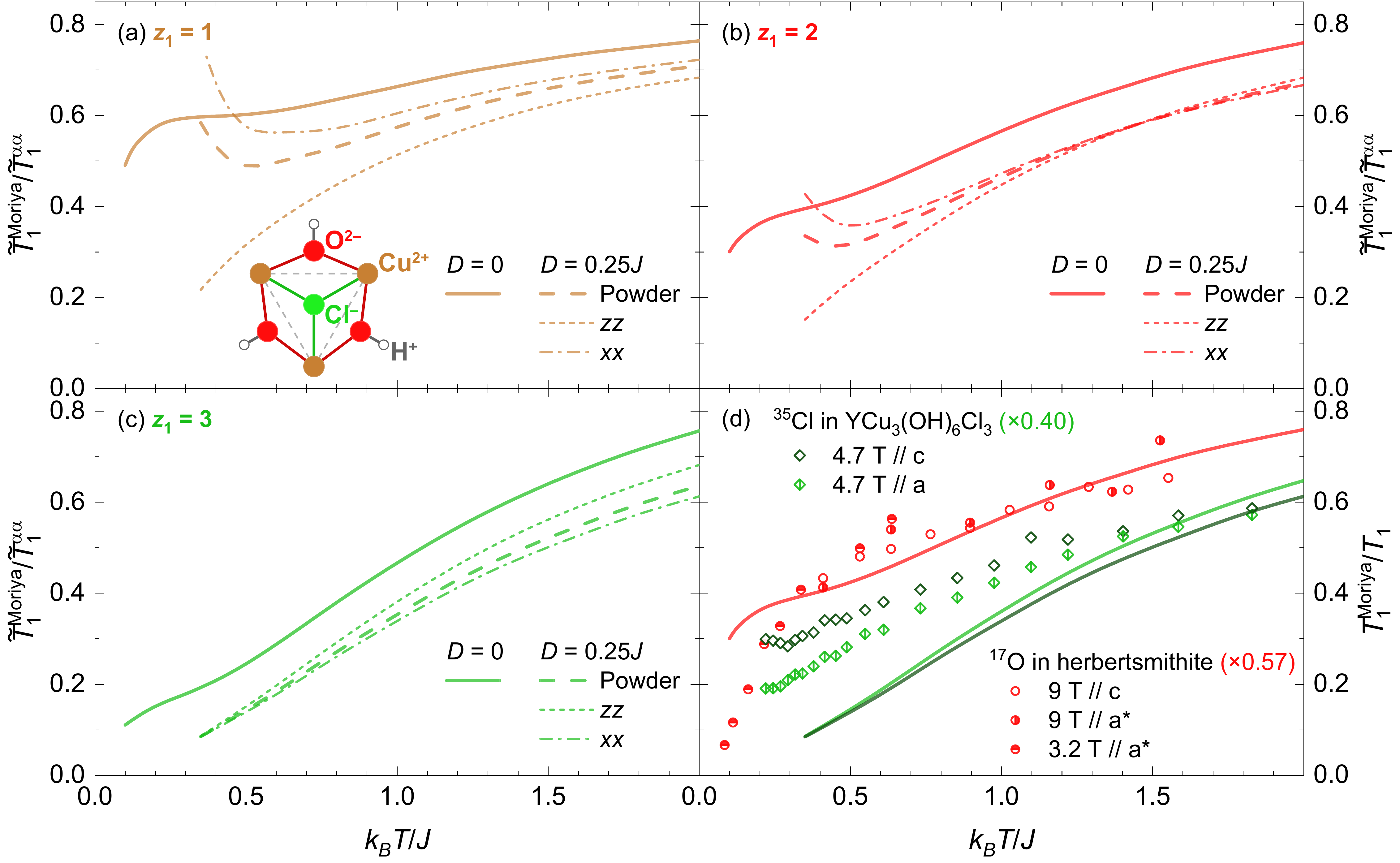}
\caption{Directional spin-lattice relaxation rate contributions $1/\widetilde{T}_1^{\alpha\alpha}$ [\cref{eqs_t1_chiral_fs,table_NMR_fc}] normalized by Moriya's Gaussian approximation $1/\widetilde{T}_1^\mathrm{Moriya}$ [\cref{eq_moriya_directional}] for nuclei coupled to $z_1$ electron spins where (a) $z_1 = 1$ (\ch{^{63{,}65}Cu}-type), (b) $z_1 = 2$ (\ch{^{17}O}-type), and (c) $z_1 = 3$ (\ch{^{35}Cl}-type nuclei) calculated for $D = 0$ and $D = 0.25 J$ on the $N = 30$ lattice.
Shown are the $\alpha = x,y$ component ($xx$), the $\alpha = z$ component ($zz$), and a powder average of both given by $1/\widetilde{T}_1^\mathrm{powder} = (1/3) \sum_\alpha 1/\widetilde{T}_1^{\alpha\alpha}$.
Representative nuclei in herbertsmithite and similar materials are shown on the inset in panel (a).
(d) Symbols show the \ch{^{17}O} NMR spin-lattice relaxation rate $1/T_1$ of herbertsmithite with $J/k_B = \SI{190}{K}$ and $z_1 = 2$ from Ref.~\cite{fu15} (green) and the \ch{^{35}Cl} NMR spin-lattice relaxation rate of \ycuFormula with $J/k_B = \SI{82}{K}$ \cite{arh20} and $z_1 = 3$ from Ref.~\cite{arh20b} (red), both normalized by \cref{eq_moriya}. 
These values are further uniformly rescaled by a factor $0.57$ in the case of herbertsmithite and $0.40$ in the case of \ycuFormula.  These are compared to FTLM results at $D = 0$ and $D= 0.25 J$ on the $N = 30$ lattice, with the curves taking appropriate averages over relevant directions $\alpha$ and chiral form factors $f_c$ in \cref{eqs_t1_chiral_fs,eq_moriya}.}
\label{fig12}
\label[pluralfigure]{figs12}
\end{figure*}

In \cref{figs12}(a--c) we show the impact of different $z_1$ on $1/\widetilde{T}_1^{\alpha\alpha}$ in more detail. 
We consider the $\alpha = x,y$ component due to spin fluctuations within the kagome plane, the $\alpha = z$ component due to out-of-plane spin fluctuations, and a powder average of both, for both zero and large $D = 0.25 J$ on the
$N = 30$ lattice. 
We present our results normalized to Moriya's Gaussian approximation \cite{moriya56} for the KLHM where $\widetilde{S}_{L c}^{\alpha\alpha,\mathrm{Moriya}}(0) = S_L^\mathrm{Moriya}(0)$ for all $c$ and $\alpha$ [\cref{eq_moriya_SL}], yielding 
\begin{equation}
\frac{1}{\widetilde{T}_1^\mathrm{Moriya}} = \frac{\hbar z_1}{8 \sqrt{\pi} J} .
\label{eq_moriya_directional}
\end{equation}
Firstly, in the $z_1 = 1$ case [\cref{fig12}(a)], where each nucleus is coupled to a single magnetic ion, 
we have $f_0 = f_{\pm 1} = 1$ (\cref{table_NMR_fc}), i.e.  all chiralities contribute equally, 
and  $1/\widetilde{T}_1^{\alpha\alpha} = S_L^{\alpha\alpha}(\omega_0 \approx 0)$ [\cref{eq_sl}].
In the intermediate case of $z_1 = 2$ [e.g.~when each nucleus is coupled equally to two spins on an exchange bond; see inset in \cref{fig12}(a)] 
we have $f_0 > f_{\pm 1} > 0$, where again all chiralities contribute to $1/\widetilde{T}_1^{\alpha\alpha}$ 
but chiral the $c = \pm 1$ contributions are suppressed compared to the $c = 0$ contribution [\cref{fig12}(b)]. 
Finally, in the case of $z_1 = 3$ 
[e.g.~when each nucleus is positioned symmetrically at or above the center of a KL triangle; see inset in \cref{fig12}(a)] 
we have $f_{\pm 1} = 0$, so that the $c = \pm 1$ fluctuations are completely filtered out, 
and just the $c = 0$ chiral LSF [\cref{figs10}(b,d)] contribute, 
resulting in a nearly isotropic $1/\widetilde{T}_1^{\alpha\alpha} = 3 \widetilde{S}_{L0}^{\alpha\alpha}(\omega_0 \approx 0)$ steadily increasing with $T$ [\cref{fig12}(c)].

\subsubsection{Experiment}

First we compare our FTLM model results with NMR experiments on herbertsmithite, \herbFormula. 
Even though several experimental NMR spin relaxation studies have been carried out on this material over the years, single-crystal studies at $k_B T>0.1 J$ relevant for comparison with our model calculations are rare. 
In \cref{fig12}(d) we summarize the \ch{^{17}O} NMR $1/T_1$ results from Ref.~\cite{fu15} measured in the direction orthogonal to the kagome planes ($c$-axis) and within the kagome planes ($a^*$-axis).
The appropriate components of the hyperfine coupling tensors $(A_a, A_{a^*},A_c) = (3.3 g_a, 4.3 g_a, 3.6 g_c) \, \si{\tesla}$ are taken from Ref.~\cite{sherman16}.
Here $g_a=2.14$ and $g_c=2.25$ are in-plane and out-of-plane components, respectively, of the \ch{Cu^{2+}} $g$-factor tensor at high-$T$ \cite{zorko17}. 
As the oxygen nuclei are positioned symmetrically with respect to two neighboring magnetic \ch{Cu^{2+}} ions [inset in \cref{fig12}(a)], we have $z_1 = 2$ and the corresponding chiral form factors are $f_0 = 4$ and $f_{\pm 1}=1$ (\cref{table_NMR_fc}).
To compare our calculations with experiment, we normalize all $1/T_1$ values to Gaussian approximation \cite{moriya56}
\begin{equation}
\frac{1}{T_1^\mathrm{Moriya}} = \frac{\sqrt{\pi} \gamma_n^2 \hbar z_1}{8 J} \sum_\alpha A_\alpha^2 ( 1 - \hat{B}_\alpha^2 ) ,
\label{eq_moriya}
\end{equation}
which can be obtained by inserting the directional $1/\widetilde{T}_1^\mathrm{Moriya}$ [\cref{eq_moriya_directional}] into the full $1/T_1$ [\cref{eqs_t1_chiral_fs}].
Like in the INS analysis in \cref{sec5a} we compare experimental results with FTLM calculations for $D = 0$, as the effect of the DM interaction on the chiral LSF at the experimentally determined value of $D=(0.04$--$0.08)J$ \cite{zorko08,el10} is very small for all directions and chiralities (see the $D = 0.05$ curves in \cref{fig10}). 
The experimental $1/T_1$ along the two crystallographic directions indeed coincide when normalized by \cref{eq_moriya}, 
and their graduate decrease with lowering $T$  nicely follows the theoretical prediction down to $k_B T \approx 0.3 J$. The downturn of the experimental \ch{^{17}O} NMR $1/T_1$ below this temperature, 
which ultimately leads to $1/T_1 \propto T^{0.84}$ below $k_B T \approx 0.05 J$ \cite{khuntia20}, 
also seems to be qualitatively supported by our model calculations, where in the latter the downturn is the signature of a 
quite robust triplet gap $\Delta_t >0$ in the considered $D = 0$ model system.

The second experimental example is the novel KL material \ycuFormula, which, like herbertsmithite, has a nearest-neighbor Heisenberg exchange coupling $J/k_B=\SI{82}{\kelvin}$ that is by far the dominant isotropic magnetic interaction \cite{arh20}. 
However, unlike herbertsmithite, this material enters a chiral $\ang{120}$ AFM LRO GS at $k_B T_N = 0.15 J$ \cite{zorko191,zorko192,barthelemy19}, which is attributed to a sizable out-of-plane DM interaction $D = 0.25 J$ \cite{arh20}. 
\ch{^{35}Cl} NMR $1/T_1$ results from Ref.~\cite{arh20b} measured in the direction orthogonal to the kagome planes ($c$-axis) and within the kagome planes ($a$-axis), 
on one of the two chlorine crystallographic sites, 
are shown in \cref{fig12}(d). 
The chosen \ch{^{35}Cl} site is coupled symmetrically with all three \ch{Cu^{2+}} spins on a given KL spin triangle [inset in \cref{fig12}(a)], 
similar to chlorine sites in herbertsmithite. 
The appropriate components of the hyperfine coupling tensor to a single electron spin are $A_{a} = A_c = \SI{0.28}{\tesla}$. 
As evident by model calculations for a nucleus in such a symmetric $z_1 = 3$ position [\cref{fig12}(c)], the anisotropy of the measured $1/T_1$ is minimal, suggesting that highly anisotropic chiral $c = \pm 1$ local spin fluctuations [\cref{figs10}(a,c)] are indeed highly suppressed at the \ch{^{35}Cl} site, 
broadly consistent with the expected chiral form factors (\cref{table_NMR_fc}). 
Nevertheless, even though the theoretically predicted trend of decreasing $1/T_1$ with lowering $T$ is followed by experiment, the experimentally-observed decrease is less pronounced [\cref{fig12}(d)]. 
As the $c = 0$ chiral LSF, which should represent the only contribution to $1/T_1$ according to \cref{table_NMR_fc}, is expected to nearly vanish at low $T$ [\cref{figs10}(b,d)], while the experimental $1/T_1$ does not, this suggests that a remnant $c = \pm 1$ contribution must still affect the experimental $1/T_1$ to a certain extent. 
This could be a telltale sign of reduced local threefold rotational symmetry in \ycuFormula, similar to the recently discovered symmetry reduction in herbertsmithite \cite{zorko17}. 

Finally, we note that the experimental $1/T_1$ results on both herbertsmithite and \ycuFormula need to be rescaled by factors of $0.57$ and $0.40$, respectively, to achieve a quantitative match with model calculations [\cref{fig12}(d)]. 
This might be partly attributed to uncertainty in experimental parameters such as the hyperfine coupling constants. 
A further source of uncertainty is also the spin diffusion contribution, i.e.~the contribution from the $q \approx 0$, $c=0$ spin fluctuations, which we omit as mentioned in \cref{sec3a}. In generic 2D systems at $\omega_0 \to 0$ this contribution might even be singular \cite{sokol93}. 
%and omitted as mentioned in \cref{sec3a}. On the other hand,
%properly treated such contribution  can even be singular in infinite 2D system in $\omega_0 \to 0$ limit \cite{sokol93}.  
Nevertheless, in systems with strong AFM fluctuations, 
like cuprates \cite{imai93} and KLHM materials, 
the spin diffusion contribution to $1/T_1$ is generally considered to be relatively small and sizable only at high $T$. 
Still, it might contribute a relevant quantitative correction to the calculated $1/T_1$.

\section{Summary and Outlook}

Our comprehensive numerical study of the dynamical spin correlations of the KL AFM via 
FTLM calculations has led to several pertinent findings. 
By separating the chiral correlations ($c = \pm 1$) from non-chiral ones ($c = 0$), we have shown that former dominate the low-energy dynamics even of the isotropic $D = 0$ KLHM (\cref{fig2,fig3}). 
These corresponds to fluctuations of the uniform ($q = 0$) $\ang{120}$ AFM order parameter, leading to a pronounced low-energy response in the DSF $S({\bf q},\omega)$ at the M point of the extended BZ. 
The dominant chiral DSF features a nontrivial frequency dependence characterized by a double-maximum structure that persists up to $k_B T \approx J$ (\cref{fig4}), even though the lower-energy peak corresponds to energies of only around $0.3 J$.  As a direct consequence of this low-energy peak, the d.c.~susceptibility $\chi_0({\bf q})$ exhibits a pronounced peak at the M point at low $T$ (\cref{fig6}). 
In clear contrast, the equal-time $S({\bf q})$, which sums over all energies, exhibit a pronounced region of high intensity that is spread out around the whole extended BZ boundary, remains stable up to high $k_B T \sim 2J$, and has apparent 
weak maxima in the corner K points of the extended BZ (\cref{fig5}).

Allowing for finite DM interactions perpendicular to the kagome plane makes the chiral DSF anisotropic (\cref{fig8}). 
Such magnetic anisotropy mainly affects the $q = 0$, $c = \pm 1$ chiral $\ang{120}$ AFM fluctuations, which soften at the quantum critical point $D_c \approx 0.1 J$. 
The corresponding out-of-plane chiral DSF response spectra $\widetilde{S}_{\pm 1}^{zz}(q=0,\omega)$ become more coherent with increasing $D$ with an increase in an effective out-of-plane spin triplet gap [\cref{fig8}(a)], while in-plane chiral DSF spectra $\widetilde{S}_{\pm 1}^{xx}(q=0,\omega)$ show enhanced low-energy fluctuations and longer-range correlations [\cref{figs8}(b) and \labelcref{fig9}]. 
The change in local (i.e.~integrated over ${\bf q}$) spin fluctuations, which are highly relevant for local-probe experiments like NMR, from the isotropic $D=0$ case is also dominated by chiral $c = \pm 1$ fluctuations (\cref{fig10}). 

All of the observed characteristic features of the KL antiferromagnet DSF can also be probed experimentally via INS and NMR spin-lattice relaxation measurements. 
We critically compare our results to two most relevant examples of the nearest-neighbor KL materials, the archetypal herbertsmithite and the novel KL material \ycuFormula. 
The former possesses rather small DM magnetic anisotropy and lacks LRO down to the lowest experimentally accessible temperatures, while the latter is characterized by a much larger DM anisotropy and chiral $\ang{120}$ LRO at low $T$. 
Single-crystal KL INS measurements with the required ${\bf q}$-space resolution are so far only available for herbertsmithite. 
These measurements indeed show a broad low-energy peak \cite{han12} at energies that are entirely consistent with the lower-energy, $0.3 J$ peak that we find numerically.   Furthermore, our model calculations also convincingly reproduce the variation of $S({\bf q},\omega)$ measured along the $(-2,1+K,0)$ ${\bf q}$-cut in Ref.~\cite{han12} (\cref{fig11}). 
%The experimental INS peak at the edge of the extended BZ is well reproduced even in quantitative detail, 
%both concerning its position as well as in its width. 
We find that the peak is considerably narrower than predicted by a simple singlet-dimer toy model \cite{han12}, which indicates that chiral AFM fluctuations in the KLHM have a finite low-$T$ correlation length $\xi > 1$.

Furthermore, \ch{^{17}O} NMR spin-lattice relaxation rate $1/T_1$ measurements on herbertsmithite \cite{fu15} are reasonably reproduced by model calculations [\cref{fig12}(d)], showing that the effect of small DM interactions that are present in this compound on dynamical spin correlations is indeed small. 
The experimental $T$-dependence is well consistent with numerical result, in particular at $k_B T > 0.3 J$. 
The observed variation predominantly reflects the evolution of the non-chiral ($c = 0$) fluctuations, as the contribution of the chiral ($c = \pm 1$) fluctuations is partly filtered out on the symmetric position of the \ch{^{17}O} nuclei.
The situation is even more extreme in the case of \ch{^{35}Cl} NMR spin-lattice relaxation rate $1/T_1$ measurements on \ycuFormula \cite{arh20b},  where chiral $c = \pm 1$ fluctuations should be completely filtered out due to rotational symmetry at the nuclear site.  Indeed, we observe almost no anisotropy in experimental $1/T_1$, however the experiment notably deviates from theory at low $T$, suggesting that the chiral contribution might
still contribute, likely due to reduced local rotational symmetry. 

Our study has demonstrated that detailed knowledge of the dynamical spin structure factor of KL AFM at $T>0$  can indeed provide invaluable insight into the nature of its low-energy spin excitations and represent
a link to numerous theoretical studies  of the ground state of this enigmatic model.
Especially intriguing is the robust, yet hitherto underappreciated, chiral nature of the dominant spin fluctuations. 
The scope of our results is further extended by the inclusion of experimentally highly-relevant DM interactions with nontrivial consequences. 
We have furthermore demonstrated that our unbiased state-of-the-art numerical calculations provide a 
reliable basis upon which past and future experiments on kagome materials can be judged and interpreted.

%--------------------------------------------------------------------------------

\acknowledgments{We acknowledge the financial support of the Slovenian Research Agency through programs P1-0044 and P1-0125, and projects N1-0088, Z1-1852, N1-0148, and J1-2461.}
%================================================================================
%\bibliography{manukago}
%
\end{document}